\documentclass[5p,times,authoryear,onecolumn]{elsarticle}

\usepackage{enumerate}

\usepackage{comment}

\usepackage{amsmath}

\usepackage{xurl}
\usepackage[nolist]{acronym}

\usepackage{xcolor}

\usepackage{amssymb}
\usepackage{scalerel}
\usepackage{stackengine}
\usepackage{lineno}

\usepackage{comment}

\usepackage[figuresright]{rotating}
\usepackage{tablefootnote}
\usepackage{amsmath, amssymb}
\usepackage{tikz}
\usepackage{tikz-qtree}
\usepackage{listings}
\usepackage{float, placeins}
\usepackage{caption}
\usepackage{subcaption}
\usepackage{tabularx} % in the preamble

\definecolor{col1}{rgb}{0.58,0,0.82}
\definecolor{ashgrey}{rgb}{0.7, 0.75, 0.71}
\definecolor{brilliantlavender}{rgb}{0.96, 0.73, 1.0}
\definecolor{ceil}{rgb}{0.57, 0.63, 0.81}
\definecolor{darkcyan}{rgb}{0.0, 0.55, 0.55}
\definecolor{celestialblue}{rgb}{0.29, 0.59, 0.82}
\definecolor{darkpastelpurple}{rgb}{0.59, 0.44, 0.84}
\definecolor{darkbyzantium}{rgb}{0.36, 0.22, 0.33}
\definecolor{darkolivegreen}{rgb}{0.33, 0.42, 0.18}
\colorlet{punct}{red!60!black}
\definecolor{background}{HTML}{EEEEEE}
\definecolor{delim}{RGB}{20,105,176}
\colorlet{numb}{magenta!60!black}

\definecolor{verylightgray}{rgb}{.97,.97,.97}
\definecolor{darkgreen}{rgb}{.0,.235,.0}

\lstdefinelanguage{circom}{
        columns=fullflexible,
	keywords=[1]{assert, do, else, false, function, if, length, public, true, while, for}, % generic keywords including crypto operations
	keywordstyle=[1]\color{blue}\bfseries,
	keywords=[2]{var, signal},	% types; money and time units
	keywordstyle=[2]\color{teal}\bfseries,
	keywords=[3]{include, template, input, output, private, main, component},	% environment variables
	keywordstyle=[3]\color{violet}\bfseries,
	identifierstyle=\color{black},
	sensitive=false,
	comment=[l]{//},
	morecomment=[s]{/*}{*/},
	commentstyle=\color{gray}\ttfamily,
	stringstyle=\color{darkgreen}\ttfamily,
	morestring=[b]',
	morestring=[b]"
}
\lstdefinelanguage{json}{
    columns=fullflexible,
    basicstyle=\scriptsize\ttfamily,
    numbers=left,
    numberstyle=\tiny,
    stepnumber=1,
    numbersep=5pt,       
    numberstyle=\tiny\color{gray},
    showstringspaces=false,
    breaklines=true,
    frame=lines,
    literate=
     *{0}{{{\color{numb}0}}}{1}
      {1}{{{\color{numb}1}}}{1}
      {2}{{{\color{numb}2}}}{1}
      {3}{{{\color{numb}3}}}{1}
      {4}{{{\color{numb}4}}}{1}
      {5}{{{\color{numb}5}}}{1}
      {6}{{{\color{numb}6}}}{1}
      {7}{{{\color{numb}7}}}{1}
      {8}{{{\color{numb}8}}}{1}
      {9}{{{\color{numb}9}}}{1}
      {:}{{{\color{punct}{:}}}}{1}
      {,}{{{\color{punct}{,}}}}{1}
      {\{}{{{\color{delim}{\{}}}}{1}
      {\}}{{{\color{delim}{\}}}}}{1}
      {[}{{{\color{delim}{[}}}}{1}
      {]}{{{\color{delim}{]}}}}{1},
}

\definecolor{lightgray}{rgb}{.9,.9,.9}
\definecolor{darkgray}{rgb}{.4,.4,.4}
\definecolor{purple}{rgb}{0.65, 0.12, 0.82}

\lstdefinelanguage{JavaScript}{
  keywords={typeof, new, true, false, catch, function, return, null, catch, switch, var, let, const, if, in, while, do, else, case, break, async, static},
  keywordstyle=\color{celestialblue}\bfseries,
  ndkeywords={class, export, boolean, throw, implements, import, this},
  ndkeywordstyle=\color{darkolivegreen}\bfseries,
  identifierstyle=\color{black},
  sensitive=false,
  comment=[l]{//},
  morecomment=[s]{/*}{*/},
  commentstyle=\color{purple}\ttfamily,
  stringstyle=\color{red}\ttfamily,
  morestring=[b]',
  morestring=[b]"
}

\usepackage{xspace}
\usepackage{relsize}
\newcommand{\Rplus}{\protect\hspace{-.1em}\protect\raisebox{.35ex}{\smaller{\smaller\textbf{+}}}}
\newcommand{\Cpp}{\mbox{C\Rplus\Rplus}\xspace}

\usepackage[hidelinks]{hyperref}

\begin{document}

\onecolumn

\begin{frontmatter}




\title{Bringing data minimization to digital wallets at scale \\ with general-purpose zero-knowledge proofs}


\author[label1,label2]{Matthias Babel}
\author[label2,label3]{Johannes Sedlmeir\corref{cor1}}
\ead{johannes.sedlmeir@uni.lu}
\cortext[cor1]{Corresponding author:}
\address[label1]{Branch Business \& Information Systems Engineering of the Fraunhofer FIT, Bayreuth, Germany}
\address[label2]{FIM Research Center, University of Bayreuth, Germany}
\address[label3]{Interdisciplinary Centre for Security, Reliability and Trust, University of Luxembourg, Luxembourg}

\acrodefplural{snark}[zk-SNARKs]{zero-knowledge succinct non-interactive arguments of knowledge}
\acrodefplural{stark}[zk-STARKs]{zero-knowledge scalable transparent arguments of knowledge}

\begin{acronym}
\acro{aca-py}[ACA-Py]{Hyperledger Aries cloudagent in Python}
\acro{bbs}[BBS]{Boneh-Boyen-Shachum}
\acro{ca}[CA]{certificate authority}
\acro{cl}[CL]{Camenisch-Lysyanskaya}
\acro{cli}[CLI]{command line interface}
\acro{crl}[CRL]{certificate revocation list}
\acro{cpu}[CPU]{central processing unit}
\acro{crs}[CRS]{common reference string}
\acro{defi}[DeFi]{decentralized finance}
\acro{dsl}[DSL]{domain-specific language}
\acro{ecdsa}[ECDSA]{elliptic curve digital signature algorithm}
\acro{eddsa}[EdDSA]{Edwards-curve digital signature algorithm}
\acro{eidas}[eIDAS]{electronic identification and trust services}
\acro{eu}[EU]{European Union}
\acro{ev}[EV]{extended validation certificate}
\acro{gdpr}[GDPR]{general data protection regulation}
\acro{irma}[IRMA]{``I reveal my attributes''}
\acro{jws}[JWS]{JSON web signatures}
\acro{jwt}[JWT]{JSON web token}
\acro{mitm}[MITM]{man-in-the-middle}
\acro{mpc}[MPC]{multi-party computation}
\acro{nfc}[NFC]{near-field communication}
\acro{ocsp}[OCSP]{online certificate status protocol}
\acro{pki}[PKI]{public key infrastructure}
%\acro{plonk}[PlonK]{permutations over Lagrange-bases for oecumenical noninteractive arguments of knowledge}
\acro{pkc}[PKC]{public-key cryptosystem}
\acro{pk}[PK]{public key}
\acro{qap}[QAP]{quadratic arithmetic program}
\acro{qwac}[QWAC]{qualified website authentication certificate}
\acro{ram}[RAM]{random-access memory}
\acro{r1cs}[R1CS]{rank one constraint system}
\acro{rsa}[RSA]{Rivest-Shamir-Adleman}
\acro{sha256}[SHA256]{secure hash algorithm~256 }
\acro{sk}[SK]{secret key}
\acro{snark}[zk-SNARK]{zero-knowledge succinct non-interactive argument of knowledge}
\acro{stark}[zk-STARK]{zero-knowledge scalable transparent argument of knowledge}
\acro{ssd}[SSD]{solid state drive}
\acro{ssi}[SSI]{self-sovereign identity}
\acro{ssl}[SSL]{secure sockets layer}
\acro{tee}[TEE]{trusted execution environment}
\acro{tls}[TLS]{transport layer security}
\acro{url}[URL]{uniform resource locator}
%\acro{vc}[VC]{verifiable credential}
\acro{von}[VON]{verifiable organizations network}
\acro{vp}[VP]{verifiable presentation}
\acro{w3c}[W3C]{world wide web consortium}
\acro{wasm}[WASM]{WebAssembly}
\acro{zkp}[ZKP]{zero-knowledge proof}
\end{acronym}

\begin{abstract}
%253 words
Today, digital identity management for individuals is either inconvenient and error-prone or creates undesirable lock-in effects and violates privacy expectations. These shortcomings inhibit the digital transformation in general and also make existing digital identity management approaches incompatible with emerging blockchain-based applications. ``Decentralized'' or ``self-sovereign'' identity aims to offer a solution by providing individuals with convenient digital wallet applications to manage cryptographic keys and machine-verifiable attestations on their edge devices. However, when presented to relying parties, these attestations typically reveal more identity attributes than required and allow for the tracking of end users' activities through unique cryptographic identifiers. Several proposals from academic research and practical solutions exist to reduce or avoid such excessive information disclosure; ranging from simple selective disclosure techniques to data-minimizing anonymous credentials constructed with zero-knowledge proofs. In this paper, we first demonstrate that currently deployed privacy-oriented self-sovereign identity solutions based on anonymous credentials still lack essential features for large-scale deployment in regulated environments. In particular, we argue that data-minimizing certificate chaining, integration with secure elements without involving a ``super cookie'', and revocation with a sufficiently large anonymity set represent essential privacy requirements that have thus far not been implemented in large-scale pilots. We then propose to address these pressing challenges by designing anonymous credentials based on general-purpose zero-knowledge proofs in the form of zero-knowledge non-interactive arguments of knowledge (zk-SNARKs). We describe our implementation and conduct performance tests on different edge devices to illustrate that the performance of our construction is already practical. We also discuss further advantages general-purpose zero-knowledge proofs can easily provide for reducing privacy risks; e.g., by facilitating customizable predicates, data-minimized credential issuance, and ``designated verifier presentations'' that avoid the risk of breaches of verifiable personal information by relying parties. 
\end{abstract}

% \sep zero-knowledge proof


\end{frontmatter}

\vspace{10em}~
\section*{Highlights}

\begin{itemize}
    \item Decentralized digital identity (SSI) requires advanced data minimization capabilities
    \item General-purpose ZKPs facilitate scalable and flexible anonymous credentials
    \item zk-SNARKs can provide private revocation, credential chaining, and hardware binding
    \item Performance can already be considered practical for rollout on mobile phones
    \item Designated verifier zk-SNARKs address pressing issues regarding security and privacy
    %\item Generic support for predicates requires new perspectives and components 
\end{itemize}

\vspace{2em}

% \linenumbers


\section{Introduction}
\label{sec:introduction}

``The Internet was built without a way to know who and what you are connecting to''~\citep[p.~1]{cameron2005laws}. 
Owing to this absence of a standardized identity layer, there is currently a ``patchwork'' of solutions for the digital identification and authentication of individuals. 
The arguably most prominent approach involves creating an account -- including a user name and a password -- for each service users interact with on the internet~\citep{preukschat2021ssi}. 
However, many individuals struggle with managing their dozens or hundreds of user names and passwords in a secure way~\citep{bonneau2012quest}. Moreover, identity attributes are largely non-transferable, i.e., they can be used only in interactions with the service provider, website, or company that created or requested them during registration and usage~\citep{sedlmeir2022ssidps}. 
Besides the tedious task of repeatedly filling registration forms, many processes that require verifiable data from, e.g., a government-issued ID~card or a university diploma, involve additional time-consuming and costly verification-related processes like video calls~\citep{sedlmeir2021digitalidentities,lacity2022self,preukschat2021ssi}.

Identity providers in federated identity management offer end users a more convenient alternative with their single sign-on services. Besides providing a consistent (and often more secure) means of authentication, these services can store users' identity attributes and forward them to relying parties, such as service providers, on the users' request~\citep{maler2008venn}. Some identity providers offer dedicated digital verification services for critical identity-related documents like ID~cards~\citep{idnow2023}. Moreover, Apple and Google have recently started integrating selected identity documents, such as digital ID~cards and driver's licenses, in their originally payment-oriented wallet apps~\citep{verge2022googlewallet}. However, many physical documents as well as proofs of authorizations, achievements, or membership that individuals need in their daily lives are still not available in machine-verifiable form. Governments also increasingly seek for means to reduce dependencies of their digital markets on foreign corporate identity providers~\citep{codagnone2023leading,ernstberger2023sovereignty}, not least because the cross-domain aggregation of an increasing variety of identity information and use-related metadata by identity providers raises significant economic, privacy, and security risks~\citep{sedlmeir2021digitalidentities,bernabe2020aries,ermolaev2023zcommerce}. 
Already in~1984, the cryptographer David Chaum hypothesized that electronic identification in general may lead to ``sophisticated marketing techniques that rely on profiles of individuals
[...] being used to manipulate public opinion and elections''~\citep[p.~1044]{chaum1985security}. More recent works advocate for a strong involvement of socio-technical research~\citep{zuboff2019age} and in particular cryptography~\citep{rogaway2015moral} in designing digital identity management systems to tackle these growing threats for society.

The current situation is daunting when considering the growing pace of the digital transformation and users' carelessness regarding the disclosure of private identity information~\citep{alashoor2022too}.
Privacy-oriented, non-proprietary digital identity management also seems particularly important in the context of emerging blockchain-based applications owing to the inherent transparency of blockchains and the related intensified issues with data protection requirements~\citep{rieger2019GDPR,schellinger2022gdpr,sedlmeir2022transparency}. For instance, central bank digital currencies~\citep{gross2021designing}, blockchain-based access control~\citep{maesa2019blockchainaccess,wu2023blockchain} and permission management~\citep{elfaqir2020overview,liao2022blockchain} require flexible and privacy-oriented digital identity management. Similar observations hold for the Metaverse -- a combination of the Internet and augmented reality via software agents~\citep{dwivedi2022metaverse,nair2022metaverse} that also commonly builds on blockchains for managing asset ownership and exchange~\citep{leenes2007privacy}. 
%Identification is a crucial component of the Metaverse yet carries substantial privacy risks~\citep{leenes2007privacy}. 
%Challenges with sensitive personal information are even further aggravated through the measurement of additional data about individuals' devices, actions, or their environment~\citep{wang2022surveymetaverse,falchuk2018socialmetaverse,nair2022exploringmetaverse}, making data minimization in the context of digital identity particularly important.

Some common alternatives for electronic identification and authentication, such as the German~eID, implement security and data minimization through secure hardware (e.g., microcontrollers integrated in smart-cards that store citizens' identity attributes). 
This design makes use of remote attestation~\citep{camenisch2017TPM} and facilitates data minimization in the form of the selective disclosure of identity attributes~\citep{margraf2011new} and range proofs (e.g., for the verification of age requirements and non-expiration) while avoiding the leakage of unique (cryptographic) identifiers~\citep{bender2010privacy}. Smart-cards can also be used in remote interactions without a dedicated device by involving the \ac{nfc} readers integrated in modern mobile phones~\citep{poller2012electronic,margraf2011new}. Yet, smart-cards are arguably not suitable for digital-native workflows, particularly when there are many different attestations, as end users must carry them around. The security-first design of the German ID~card also does not smoothly extend to the variety of attestations with heterogeneous security, privacy, and accessibility needs reflecting the different organizations and processes users interact with in their daily lives~\citep{schellinger2022mythbusting}. On the other hand, secure hardware also generally exhibits a tradeoff between functionality and security. For instance, embedded secure elements on mobile phones can achieve the strong identity assurance required in some regulated processes~\citep{bsi2019mobileidentity} but typically only support storing cryptographic keys and some standardized operations to create digital signatures with them. Extending the use of secure elements to data minimization hence requires the cooperation of the corresponding manufacturer and -- in the case of the mobile phone -- operating system providers. Such cooperations seems challenging to implement even for a single manufacturer and selected devices, as the German government's efforts with Samsung suggest~\citep{ergo2022smarteID,bundesregierung2023antworteid}. On the other hand, the more flexible trusted execution environments in mobile phones are known to be vulnerable to sidechannel attacks~\citep{jauernig2020trusted} and, thus, cannot provide the security levels that some highly regulated workflows require~\citep{schellinger2022mythbusting,bsi2019mobileidentity,verheul2021secdsa}.

A recent approach in digital identity management aims to give users both convenience and control through empowering them to self-manage their attestations in ``digital wallet'' applications on their mobile devices~\citep{sedlmeir2021digitalidentities,cucko2021decentralizedSSI}. ``Issuers'' confirm users' identity attributes by providing them with digital certificates that carry electronic signatures. Upon request, individuals can use their digital certificates to reveal selected identity attributes to relying parties in a cryptographically verifiable way, without the need to again interact with the issuer. Governments increasingly support this decentralized or \ac{ssi} paradigm; with large-scale pilots, such as Canada's \ac{von} and the European IDunion consortium, exploring the approach~\citep{sedlmeir2022ssidps}. Moreover, the \ac{eu} is currently shaping a revision of its \ac{eidas} regulation that mandates every member state to provide its citizens with a digital wallet app that can receive, store, and present such digital attestations~\citep{rieger2022notyet,eu2022wallet}. The \ac{eidas} regulation also formalizes strict requirements in terms of security in the form of different ``levels of assurance''. Achieving the level ``high'' is required for several regulated interactions with both public and private institutions and mandates strong protections against \ac{mitm} attacks, e.g., through the use of embedded secure elements, timely means of revocation, and the mandatory certification of stakeholders that can interact with it~\citep{verheul2021secdsa}.

In its simplest form, the \ac{ssi}-based approach involves the sending the digital certificate directly to the relying party, which then verifies the corresponding digital signature. However, providing the entire certificate to the relying party reveals a significant amount of information that is not strictly necessary in the given context~\citep{hardman2020zkpsavvy, brands2000rethinking}. This unnecessary information includes identity attributes included in the certificate but not required by the relying party for the corresponding workflow. Consequently, the architecture reference framework for the European digital wallet foresees ``selective disclosure'' capabilities~\citep{ec2023arf} that only reveal the identity attributes that are explicitly required by the relying party. To give another example, the value of the digital signature on a digital certificate represents a unique cryptographic identifier that can be used to track individuals whenever they use the certificate -- an example of a certificate-specific ``super cookie''~\citep{evernym2020credentials}. The presence of such unique identifiers has recently sparked a controversial debate on the privacy aspects of the European Digital Wallet, with ongoing controversial discussions in the project's GitHub repository~\citep{eudi2023githubprivacydiscussion} and an open letter signed by more than 300~researchers as well as NGOs expressing their concerns, among other issues, about the lack of mandatory support for unlinkability guarantees in the current architecture reference framework~\citep{eIDAS2023openletter}. This paper aims to argue how such unlinkability guarantees can be implemented in practice while aligning with other regulatory and business requirements.

\textbf{Anonymous credentials} resolve the above-mentioned privacy issues arising from unique cryptographic information by enabling users to present their attestations in a data-minimal way~\citep{chaum1985security,brands2000rethinking,camenisch2001efficient,kaaniche2020privacy,kakvi2023sokanonymouscredentials}. Users reveal only selected information derived from their digital certificate that is indispensable for the respective purpose; while maintaining cryptographic verifiability. In particular, anonymous credentials aim to avoid the presence of unique cryptographic identifiers in interactions between users and relying parties and, therefore, provide unlinkability -- at least beyond linkability through the identity attributes the verifier requests. This can be achieved with \acp{zkp}, which allow a prover to convince a verifier of a statement without conveying any information apart from the statement's validity~\citep{goldwasser1989knowledge,camenisch2002design}. \acp{zkp} can be used, for instance, to confirm that a presented identity attribute is part of a certificate issued by a certain institution without having to reveal the value of the digital signature~\citep{hardman2020zkpsavvy,feulner2022ticketing}. Several digital wallets already support handling multiple of these anonymous credentials and generating the corresponding \acp{zkp}~\citep{linux2022aries,sartor2022ux,gloeckler2023ssiiam}. However, these implementations of anonymous credentials rely on academic works that involved significant effort in hand-crafting the cryptographic primitives they use~\citep{camenisch2001efficient,sudarsono2011efficient}. They were major breakthroughs at the time of their publication and allow for fast proof generation, transmission, and verification. Yet, being highly tailored to a specific set of functionalities also implies that highly specialized cryptographers need to develop novel ideas to incorporate additional features that appear, for instance, owing to regulatory constraints. Indeed, existing anonymous credential implementations typically lack both cryptographic agility and auditability~\citep{young2022real,kakvi2023sokanonymouscredentials} and hence struggle with integrating into existing digital identity infrastructures~\citep{rosenberg2022zkcreds,yeoh2023fidoac}. For instance, major \ac{ssi} projects have proclaimed the need for privacy-preserving credential chains, also called ``private delegation'' for years~\citep{aries2022chaining} (see more details in Section~\ref{subsec:credential_chaining}), and solutions were indeed found (with incrementally improving features and performance) years after the initial conceptualization of anonymous credentials~\citep{chase2006signatures,belenkiy2009randomizable,camenisch2017practical}. Yet, they still remain to be implemented in \ac{ssi} pilot projects building on anonymous credentials~\citep{aries2022chaining}.
Moreover, recent discussions have pointed out further shortcomings of anonymous credential systems built on purpose-specific \acp{zkp}, e.g., for equipping anonymous credentials with scalable revocation~\citep{young2022real} and supporting hardware binding with common mobile phones~\citep{feulner2022ticketing}.

In this paper, we detail the above-mentioned challenges, as well as further shortcomings of anonymous credentials constructed from special-purpose \acp{zkp}, in the context of practically deployed anonymous credential systems. We argue that using general-purpose verifiable computation in the form of \acp{snark} that have matured in the context of cryptocurrency privacy~\citep{benSasson2014zerocash,buterin2023blockchain,feng2019surveyblockchainprivacy} and scaling~\citep{thibault2022rollup,lavaur2023rollup,gangwal2023surveylayertwo} projects over the last decade allows addressing pressing shortcomings that have thus far inhibited the broad adoption of anonymous credentials. We thus bridge the related research streams from cryptography on the design of special-purpose \ac{zkp}-based anonymous credentials~\citep[e.g.,][]{camenisch2001efficient,sudarsono2011efficient,camenisch2017practical} and
\ac{snark}-based) anonymous credentials~\citep[e.g.,][]{kosba2015coco,delignat2016cinderella, schanzenbach2019zklaims,rosenberg2022zkcreds, maram2021CanDID,heiss2022onchain,yeoh2023fidoac} with requirements from the \ac{ssi} domain~\citep[e.g.,~][]{cucko2021decentralizedSSI,soltani2021surveyssi,sedlmeir2022ssidps,young2022real,schwalm2022eidas}. We do so grounded on our own implementation of \ac{snark}-based anonymous credentials\footnote{See \url{https://github.com/applied-crypto/heimdall}.} and experiences from our active involvement in several \ac{ssi} projects and workshops in industry and the public sector. %in Germany and Luxembourg.
We structure our work as follows. First, we give a basic understanding of \ac{ssi}, the concept of digital attestations stored in digital wallet applications, related terminology, and technical foundations of \acp{snark} in Section~\ref{sec:background}. Next, in Section~\ref{sec:related_work}, we comprehensively survey related work on anonymous credentials in cryptography research and practice. Instead of outlining the mathematical and cryptographic foundations of these constructions, we focus on discussing which requirements for broad adoption existing implementations address and which features are still missing. After that, we describe one way of implementing anonymous credentials including these lacking features with \acp{snark} (Section~\ref{sec:design}). We also conduct a performance analysis to demonstrate that this approach can be considered practical for use in mobile wallet apps as of today in Section~\ref{sec:evaluation}. In Section~\ref{sec:discussion}, we discuss further requirements of \ac{ssi} that general-purpose \acp{zkp} can address; including customizable predicates, data-minimized issuance, and designated verifier presentations. We conclude by mentioning remaining limitations and pointing towards avenues for future research in Section~\ref{sec:conclusion}. 
%The Appendix includes additional information on our credential design and code snippets from the implementation. 
\section{Background}
\label{sec:background}

\subsection{Self-sovereign identity}

The paradigm of decentralized or user-centric digital identity, also called \ac{ssi}, empowers an individual to self-manage digital attestations of their identity attributes and cryptographic keys for authentication on their edge devices (e.g., their mobile phones) in a digital wallet app~\citep{sedlmeir2022ssidps,weigl2023ssi}. Digital attestations are created by ``issuers'' -- entities such as public sector institutions, enterprises, individuals, or machines associated with a cryptographic key-pair. Issuers are considered to have a certain reputation within specific domains~\citep{soltani2021surveyssi}. For instance, a meaningful institution to issue digital national~IDs could be the public institution that currently manufactures physical ID~cards, or a country's signing certification authority (CSCA) as governed by the International Civil Aviation Organization (ICAO). Analogous to watermarks and seals on physical documents, digital attestations carry cryptographic proofs of integrity; usually a digital signature generated by the issuer~\citep{sedlmeir2021digitalidentities}. This digital signature makes the digital certificate machine-verifiable, which is why the corresponding attestations are often termed ``verifiable credentials''. 
At the same time, ``Verifiable Credentials'' refers to a nascent standard established by the \ac{w3c} to harmonize digital attestations~\citep{w3c2019vc}. Many emerging \ac{ssi} projects related to this \ac{w3c} standard build on or have built on blockchains~\citep{sedlmeir2021digitalidentities}; for instance, to provide highly available revocation registries~\citep{tesei2023blockchainrevocation,feulner2022ticketing} and decentralized and more transparent alternatives to the existing \acp{pki} that maps organizations to their public keys~\citep{schlatt2021kyc}.
Because there are also alternative, more established standards for digital certificates (see Section~\ref{sec:related_work}), we will neutrally use the term ``credential'' in the following to cover all different flavors of digitally signed and, therefore, machine-verifiable attestations.
Note that this terminology is more narrow than the one proposed by \citet{bosworth2005entities}, according to which a credential is ``used to prove an identity to a system'', i.e., a physical token or password would also qualify as a credential.
It will also become apparent that for data minimization using general-purpose \acp{zkp}, the design of the credential itself has only limited relevance~\citep[see, e.g., also][]{delignat2016cinderella}, which is why by writing ``credential'' we explicitly aim to be inclusive of anonymous credentials.

Individuals, called ``holders'', can use their credentials to conveniently disclose identity attributes to ``verifiers'', i.e., relying parties. Typically, a verifier first sends a ``proof request'' to the holder, including a random number to prevent replay attacks (see below) and asking for the disclosure of certain attributes stated in one or several of the holder's credentials. The proof request also lists a set of additional parameters (e.g., a timestamp to refer to for expiration- and revocation-related requirements) and constraints (e.g., a list of issuers that the verifier trusts for each individual identity attribute)~\citep{preukschat2021ssi,gloeckler2023ssiiam}. When receiving the proof request, the holder's digital wallet app can fully automatically search for stored credentials that include the requested attributes and that satisfy the requirements specified in the proof request~\citep{sartor2022ux}. Upon the holder's consent on revealing the attributes specified in the proof request, the wallet app then creates a cryptographic proof of the correctness of these attributes according to the respective issuer(s) and sends the attributes and the proof to the verifier~\citep{feulner2022ticketing,gloeckler2023ssiiam}. The verifier can then algorithmically check the proof and, therefore, the authenticity of the identity attributes as claimed by the holder, and subsequently use and process them for providing its service.

The process that starts with a verifier's proof request and ends with the verification of the proof that the holder created by the verifier is called ``\ac{vp}''~\citep{schlatt2021kyc,feulner2022ticketing}. The simplest and arguably least privacy-oriented \ac{vp} involves sending one or multiple credentials that include the requested identity attributes directly to the verifier. The verifier can extract the required attributes and verify the issuer's digital signature and the fulfillment of the other requirements from the proof request directly (e.g., non-expiration) for each credential. This approach implies that the verifier could forward credentials to other parties and impersonate the holder. To mitigate such simple replay attacks, this common type of \ac{vp} must include a digital signature with the holder's ``binding key'' on the random number (``challenge/nonce'') communicated by the verifier in the proof request for each credential. The binding public key is a part of the credential, whereas the holder never shares their private binding key~\citep{dutto2020pqzkpvc,chadwick2022openid}.  
%Holder binding is also essential when the corresponding binding key-pair needs special protection, for instance, because regulators demand particularly high security both regarding attacks and the potential voluntary sharing of credentials. In these cases, smart cards or smartphone-embedded secure elements can be used to generate the key-pair and protect the corresponding secret key.

While the challenge-response mechanism can fix the security issue associated with sharing the full credential, it does not prevent excessive information disclosure. A credential may include considerably more identity attributes than the verifier requested from it. Moreover, certain cryptographic data included in the credential, such as the value of the issuer's digital signature and the holder's binding public key can be considered globally unique identifiers~\citep{brands2000rethinking}. Hence, more privacy-focused approaches to \acp{vp} do not communicate the full credential to the verifier but instead only reveal selected attributes and provide blinded cryptographic evidence derived from the credential that these attributes are indeed attested by the specified issuer~\citep{hardman2020zkpsavvy,sedlmeir2022ssidps}. Such derived proofs are usually constructed using a \ac{zkp} (see Section~\ref{subsec:zkp}), and credentials that support such derived proofs are usually termed anonymous credentials~\citep{kaaniche2020privacy}. Beyond facilitating ``selective disclosure'' and hiding the digital signature and binding public key included in the credential, sometimes it is also desirable to reveal only the results of a (potentially complex) computation that uses identity attributes as parameters to the verifier. Well-known examples of such ``predicates'' or ``predicate proofs'' include set (non-)~membership proofs, for instance, to demonstrate that a credential is not included in a revocation list. Revocation is needed in case keys or credentials are lost or  stolen~\citep{kocher1998certificate}, or if the reason for eligibility ceases to exist. For example, if an employee quits their job, the employee badge that grants access to the employer's buildings or IT~resources needs to be revoked~\citep{gloeckler2023ssiiam}
. Other examples of predicates include range proofs, e.g., to show that a date of birth as recorded in a credential is more than 18~years in the past~\citep{hardman2020zkpsavvy}.

\subsection{Zero-knowledge proofs and zk-SNARKs}
\label{subsec:zkp}

\acp{zkp} are defined as ``those proofs that convey no additional knowledge other than the correctness of the proposition in question''~\citep{goldwasser1989knowledge}. They build on proofs that ensure ``soundness'' -- a guarantee for the verifier that the prover's statement is indeed correct -- with high probability and not with certainty as in common mathematical proofs~\citep{evans2023succinct}. In this model, the prover's private information -- the ``witness'' -- can be used to convince the verifier of the statement without revealing the witness. The formal definition of the zero-knowledge property involves the notion of a ``simulator'' that does not have access to the witness but nevertheless can produce transcripts which -- except for timing-related aspects that are critical to ensure soundness -- are indistinguishable from the proofs generated by a prover with access to the witness from the verifier's perspective~\citep{goldwasser1989knowledge}. A simple example of a \ac{zkp} is proving knowledge of a secret key associated with a public key with Schnorr's protocol, giving away no information that would make it easier for the verifier to find the secret key~\citep{schnorr1991efficient}. More generally, ``proofs of knowledge'' do not only convince the verifier of the correctness of a statement (which may be trivial, e.g., the existence of a secret key associated with a given public key) but additionally that the prover indeed knows such a concrete witness. This notion can be formalized using ``extractors'' that can efficiently derive a witness from having direct access to any proving algorithm that convinces the verifier with non-negligible probability~\citep{thaler2020proofs}. 

\begin{figure*}[!tb]
    \centering
    \includegraphics[page=2, width=0.5\linewidth, trim=0cm 0.8cm 19cm 1cm, clip]{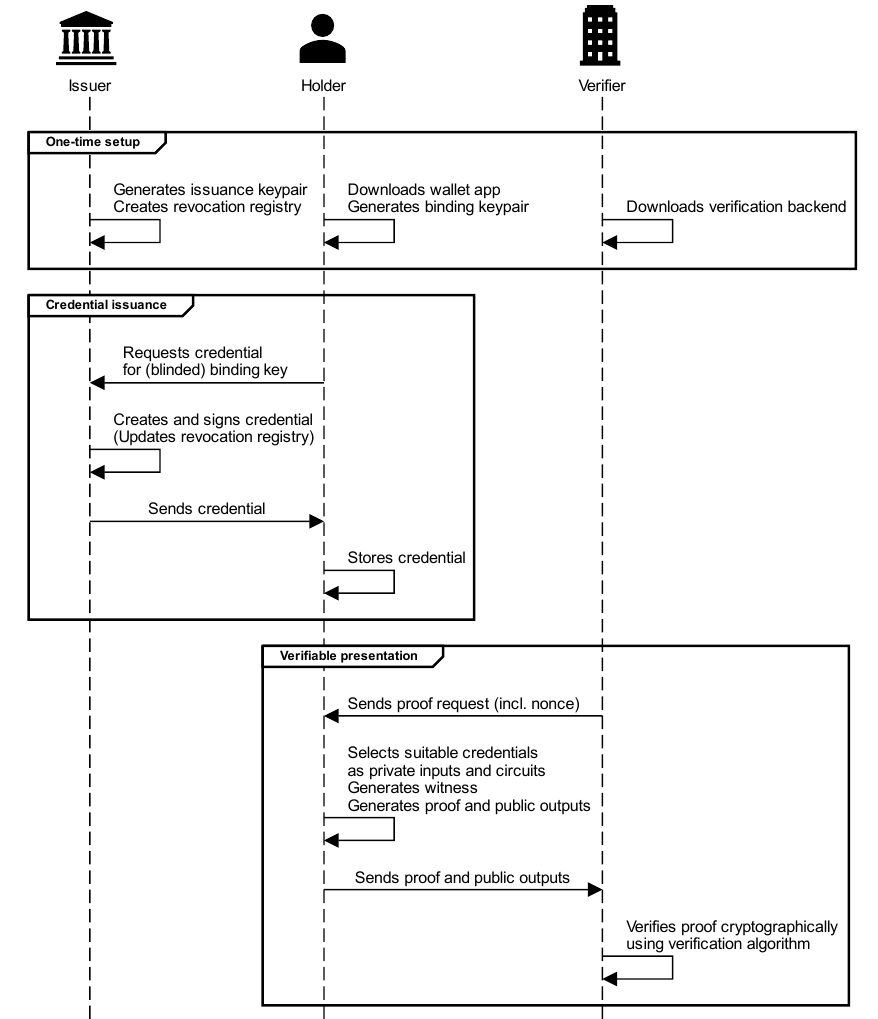}
    \caption{Issuance and verification of a credential using general-purpose \acp{zkp}.}
    \label{fig:issuance_and_verification}
\end{figure*}

A generalization of the mathematical ideas underlying Schnorr's protocol-- performing mathematical tricks in the context of the discrete logarithm problem that is assumed to be hard -- also builds the basis for anonymous credentials based on \ac{cl} signatures~\citep{camenisch2001efficient,maurer2009unifying}. (Probabilistic) proofs whose soundness relies on cryptographic hardness assumptions are more formally referred to as ``arguments'' rather than proofs~\citep{thaler2020proofs}. 
\ac{cl} signatures represent an example of hand-crafted, special-purpose \acp{zkp}. They are highly efficient in the sense that the proofs are small (several hundred bytes) and fast to create and verify (tens of milliseconds on a commodity laptop)~\citep{kakvi2023sokanonymouscredentials}. Since ``everything provable is provable in zero knowledge~\citep{ben1988everything} and there are generic compilers for transforming any algorithm that verifies a certain statement into a zero-knowledge prover and verifier,
\citet[][p.1]{camenisch2002signature} already mentioned the opportunity to create anonymous credentials ``using general-purpose zero-knowledge proofs''. However, at that time, this approach ``require[d] expensive computations beyond what is considered practical''. Indeed, creating a \ac{zkp} for the correct execution of a complex algorithm (where the witness is the correct computational trace, in particular, including all inputs and outputs) was long prohibitively computationally expensive. However, it has now become practical after almost four decades of substantial improvements in construction and silicon, for instance, in the form of \acp{snark}~\citep[e.g.,~][]{bensasson2013snarks,gennaro2013quadratic,groth2016size,parno2016pinocchio}. While the use of \acp{snark} for anonymous credentials has been advocated already early~\citep{kosba2015coco,delignat2016cinderella} (see also Section~\ref{sec:related_work}), industry use of \acp{snark} has first appeared in the context of blockchains. Initially, \acp{snark} were used there for proving still relatively simple statements to provide private payments in cryptocurrencies such as Zcash~\citep{benSasson2014zerocash}. Further improvements now also make them applicable to complex statements that facilitate, for instance, improving blockchains' transaction throughput in ``zk-rollups''~\citep{simunic2021verifiable,thibault2022rollup,gangwal2023surveylayertwo}. As operations on blockchains are one-to-many, inefficient, and expensive owing to the replicated storage and execution of transactions~\citep{bensasson2018scalablepreprint}, for both these applications, ``non-interactiveness'' and ``succinctness'' of \acp{snark} is essential. The former property describes that a single message by the prover convinces every verifier; the latter that proof size is very small and the computational complexity of proof verification is very low. (In fact, many zk-rollups leverage SNARKs that are not zero-knowledge.)

General-purpose \acp{zkp} and in particular \acp{snark} hence introduce a novel paradigm of \acp{vp}: Instead of sending credentials to the verifier, who then runs the cryptographic verification algorithm, and instead of presenting highly specialized mathematical tricks using the information in the credential in popular anonymous credential constructions, the \emph{holder} runs the verification algorithm on their device using the locally stored credential(s), and only sends the verification result and selected attributes or predicates that need to be disclosed to the verifier~\citep{kosba2015coco,delignat2016cinderella}. To allow the verifier to trust in this verification result, the holder also creates a \ac{zkp} that certifies the correct execution of the verification program and sends it to the verifier, yet without sharing any details about the inputs and intermediary results of running the credential verification algorithm. In other words, a \acp{zkp} can convince the verifier that the verification algorithm that the holder ran terminated with the specified result. As such, the ``statement'' to be proven in zero-knowledge can be of the form ``the holder knows a credential that was indeed issued by an institution with public key 0x1234 (i.e., digitally signed by the corresponding secret key), and the holder knows the secret key associated with the binding public key. Moreover, the credential is neither expired nor revoked, and the first name according to the credential is Alice''. Figure~\ref{fig:issuance_and_verification} illustrates the overall flow of issuing a credential and performing a \ac{vp} with a general-purpose \ac{zkp}, which hardly differs from \acp{vp} building on special-purpose \acp{zkp} (see, e.g., \citet{schlatt2021kyc}).

A challenge related to the use of \acp{snark} besides the computationally intensive proof generation (see Section~\ref{sec:evaluation}) is that the first practical variants required an initial computationally and memory intensive preprocessing process called ``trusted setup''~\citep{bensasson2013snarks,parno2016pinocchio}. In this trusted setup, a \ac{crs} that is required for generating and verifying \acp{zkp} is computed. It can be conducted in a \ac{mpc}. If at least one party that participates in the creation of the \ac{crs} is honest, provers cannot efficiently create fake proofs. Privacy guarantees are even unconditional~\citep{fuchsbauer2018subversion}. For blockchain applications where decentralization and ``trustlessness'' are critical, the \ac{crs} is typically generated in a massive \ac{mpc} that can involve hundreds of participants~\citep{bowe2017setup}. While there are different ways to translate an algorithm into a format that allows to generate a \ac{crs} and, ultimately, proving and verification programs, with compilers being available also for C~code~\citep{parno2016pinocchio}, to date the more efficient way seems to be through \acp{dsl} such as Circom~\citep{iden32022circom} or ZoKrates~\citep{eberhardt2018zokrates}. A frequently used representation of a statement corresponding to the correct execution of a program is the \ac{r1cs}. It specifies all the constraints that a valid computational trace to the program must satisfy. In this representation, every constraint in the execution of the program can be represented by a quadratic expression. A convenient proxy for the complexity of the statement with regard to proving effort is the number of such ``non-linear \ac{r1cs} constraints'', which for the popular ``Groth16'' proof system that underlies our prototype~\citep{groth2016size} roughly corresponds to the number of multiplications and is proportional to the proving time.
One disadvantage of the Groth16 proof system is that the trusted setup is program-specific, i.e., every update of the algorithm for which correct execution needs to be proved requires a new trusted setup. More recent flavors of \acp{snark} like Plonk~\citep{gabizon2019plonk} do not require a circuit-specific trusted setup but instead only need to create one ``universal'' \ac{crs} that can be used for all algorithms up to a certain complexity threshold (similar to the number of \ac{r1cs} constraints). The higher flexibility of universal \acp{snark} typically comes with some trade-offs, such as larger proof sizes and higher verification complexity. A remarkable alternative is given by transparent \acp{snark}, such as \acp{stark}, which replace the \emph{trusted} setup by a transparent one that is still computationally intensive but is not vulnerable to the collusion of all its participants. Transparent \acp{snark} tend to involve substantially larger proof sizes of tens to hundreds of kilobytes for typical programs (scaling with $\log^2(N)$ where $N$ represents the complexity of the program)~\citep{bensasson2018scalablepreprint}. Yet, for a bilateral interaction between a prover and a verifier, this proof size can still be considered moderate.

\begin{figure}[!tb]
    \centering
    \includegraphics[page=1, width=\linewidth, trim=0cm 1cm 13cm 0cm, clip]{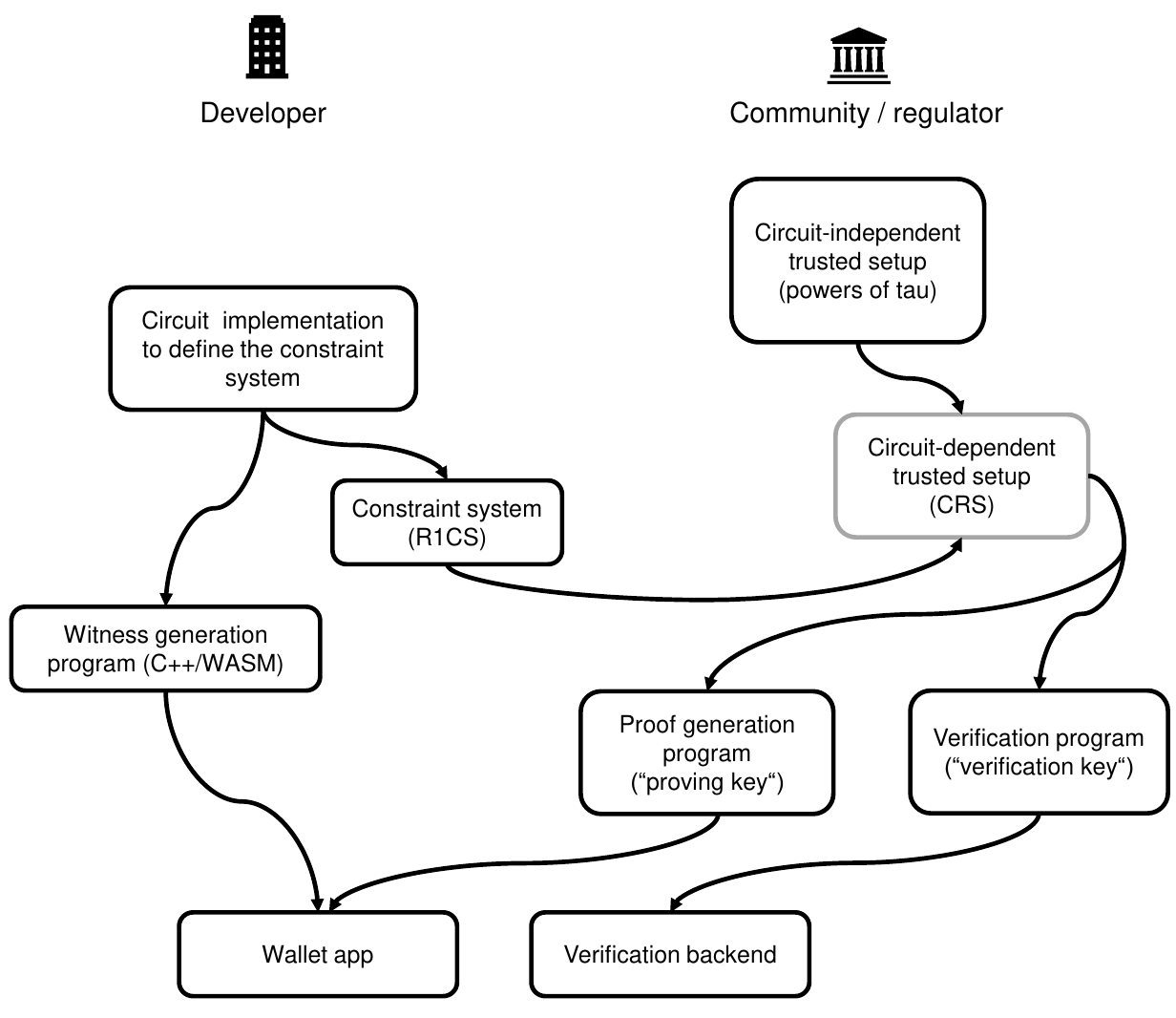}
    \caption{Example workflow for generating the \ac{snark} proving and verification program in the case of the Groth16 proof system. These programs need to be integrated into the wallet app and verification backend, respectively.} %Note that the circuit-specific trusted setup is not necessary for universal \acp{snark}. Transparent \acp{snark} do not involve any trusted setup. However, to achieve succinctness, both universal \ac{snark} and transparent \acp{snark} still involve a computationally intensive preprocessing step.}
    \label{fig:setup}
\end{figure}

Figure~\ref{fig:setup} features an overview of the steps required to set up an anonymous credential system using the Groth16 proof system.
We implemented all \acp{snark} using the \ac{dsl} Circom (``Circuit compiler'')~\citep{iden32022circom}.
Circom utilizes a finite field, with the number of field elements being a 254-bit prime number. We will often use 253-bit integers in the following to represent identity attributes, as we can embed these (via a canonical injective map) into the 254-bit prime field. Non-integer values are not natively supported, which is why additional logic for handling Strings or Floats needs to be provided (see Section~\ref{subsec:encoding}). Circuits implemented in Circom involve ``Signals'' to explicitly define constraints. One can directly use quadratic constraints or build on components -- instantiations of previously defined ``Templates'', comparable to libraries. 
%Programming them is restricted, on behalf of the underlying \ac{qap}, to use only quadratic calculations inside one Signal. 
Both signals and components must be assigned once and are immutable. For this reason, some calculations have to be split into multiple sub-calculations (see, for instance, Figure~\ref{code:cubic} for the simple case of a cubic expression). There is also no native support for branching operations, such as \texttt{if}, \texttt{break}, or \texttt{continue} statements when branching depends on factors that are not known at compile time of the proving and verification algorithm, i.e., when they depend on input signals. On the other hand, Circom already provides libraries that implement comparators (e.g., \texttt{LowerThan}), conversions between numbers and their binary representation (\texttt{Bitify}), hash functions such as \texttt{Poseidon} (see Figure~\ref{code:pre-image}) and \acs{sha256}, and signature mechanisms such as the \ac{eddsa} on the Baby JubJub elliptic curve~\citep{iden32022circomlib}. Projects that build on Circom use these building blocks to implement more advanced or complex primitives, e.g., for the verification of Merkle proofs~\citep{github2022rollup} and digital signatures with the
\ac{ecdsa}~\citep{github2022ecdsa, github2022ecdsaoptimized}.

Circom compiles a circuit into a constraint system and a witness generation program, which derives an assignment for all (intermediary) signals from the input signals. SnarkJS~\citep{iden32022snarkjs} provides means to generate a \ac{crs} and derive proving and verification programs from the constraint system, described via \ac{r1cs}. By default, all signals remain private, and only explicitly defined output signals -- called ``public outputs'' -- are revealed to the verifier. The proving program includes two parts; witness (i.e., computational trace as described by the individual Signals) generation in \ac{wasm} and cryptographic proof generation and proof verification in Javascript (Node.js) for the Groth16 (circuit-specific trusted setup) and Plonk (universal trusted setup) proof system. For productive use, PolygonID has developed highly optimized \Cpp and Assembly-based witness generation~\citep{polygon2023witnesscalc} and an Assembly-based prover -- ``Rapidsnark'' -- for multiple platforms, such as Intel x86\_64, Linux/macOS ARM, Android, and iOS~\citep{polygon2023rapidsnark}. Moreover, witness generation, Groth16 proof generation, and proof verification can be conducted in Rust via the ark-circom crate~\citep{gaconst2022ark-circom}, based on the \ac{wasm} files, proving, and verification keys generated from Circom and SnarkJS.

\section{Related Implementations and Academic Work}
\label{sec:related_work}

\subsection{X.509 certificates}
The X.509~standard is broadly adopted on the internet as a fundamental component of the the Hypertext Transfer Protocol Secure (HTTPS)~\citep{cooper2008internet}. 
These credentials are mostly hierarchically organized in what is called ``credential chains''. For instance, a \ac{ca}, as one of the most trustworthy institutions in the ecosystem, can create a credential that binds a company to a domain and cryptographic key-pair. 
This company, in turn, can use the binding key-pair corresponding to this attestation to issue a credential to one of its web servers. 
Lastly, the web servers can identify themselves through this credential and, thus, establish secure connections to clients.
Digital signatures are permanent; yet, sometimes, issuers realize that the reason for issuance ceases to exist prior to expiration. 
As the deletion of information can hardly be enforced, X.509~certificates hence carry a unique serial number that can be used in a \ac{vp} for checking their revocation state. The holder can interact with the issuer or the responsible \ac{ca} according to the \ac{ocsp} to get a short-lived, signed confirmation about the non-revoked state that they can attach to the certificate when presenting it~\citep{delignat2016cinderella}. Alternatively, \acp{crl} can be used; where the verifier would download a list of all revoked certificates from the issuer or \ac{ca} that the issuer defined as responsible for maintaining the \ac{crl}~\citep{cooper2008internet}.

As the holder transmits X.509 credentials entirely to the verifier, the corresponding \ac{vp} is far from data minimizing. A simple modification that does not include the attributes directly in the credential but instead only each attribute's salted hash~\citep{deSalve2022selective} or a single Merkle root~\citep[e.g.,~][]{liu2018selective, mukta2020selective} facilitates selective disclosure. In a \ac{vp}, the holder would transfer the full credential plus selected attributes, including the corresponding salt values or Merkle proofs~\citep{merkle1987digital}. Yet, sophisticated correlation attempts based on the digital signature, binding key, and serial number, which represent globally unique identifiers with high probability~\citep{brands2000rethinking}, are not prevented. Consequently, while X.509 certificates have been remarkably successful for the identification of servers on the web, they seem less suitable for the privacy-oriented digital identity management of natural persons without further modifications.

%% MDOC / mDL would also be interesting

\subsection{Hyperledger AnonCreds}
There are plenty of \ac{ssi} implementations, and arguably none of them currently occupies a major role in terms of practical adoption~\citep{sedlmeir2021digitalidentities}. The Hyperledger AnonCreds~\citep{hyperledger2022anoncreds} that build the foundation of the protocols specified by Hyperledger Aries~\citep{linux2022aries} and related implementations, such as the \ac{aca-py}~\citep{schlatt2022eprescription} and several compatible digital wallets (e.g., esatus, Lissi, and Trinsic)~\citep{sartor2022ux} are arguably among the implementations with the most sophisticated privacy functionalities. The technical backbone of Hyperledger AnonCreds goes back to work by~\citet{camenisch2001efficient}, building on purpose-specific \acp{zkp}. They enable not only selective disclosure but also hide the issuer's digital signature on the credential, as well as the binding public key. Moreover, they allow proving non-revocation with a zero-knowledge set-membership proof that does not expose a credential-specific identifier, such as a serial number or revocation~ID. Lastly, they support range proofs. Besides Hyperledger AnonCreds, \citep{camenisch2001efficient} builds the foundation of many other implementations of anonymous credentials~\citep{kakvi2023sokanonymouscredentials}, such as IBM's Identity Mixer~\citep{bichsel2009idemix} underlying the \ac{irma}~\citep{alpar2017irma} and European ARIES research project~\citep{bernabe2020aries}.
Similar features, yet with substantially higher performance and smaller proof sizes~\citep{andrea2023clvsbbs}, are provided by the approach of~\citet{sudarsono2011efficient} through moving from an \ac{rsa}-based approach to short signatures based on elliptic curve pairings, first suggested by \ac{bbs}~\citep{boneh2004short}. Related anonymous credentials designs are known as \ac{bbs}+.
Some digital wallet implementations build on this work~\citep{mattr2023bbs,dockio2022bbslegosnark}, and also the \ac{aca-py} has recently integrated \ac{bbs}+ credentials~\citep{acapy2022}. Besides these distinguished streams, there is also a variety of constructions building on slightly different special-purpose and efficient cryptographic primitives, such as blind signatures~\citep{paquin2011uprove} and attribute-based signatures~\citep{kaaniche2016abs}.
%\paragraph{Further blockchain-based anonymous credential systems}
Originally, the Hyperledger AnonCreds were strongly connected to a public permissionless blockchain network that provides a \ac{pki} and revocation registries~\citep{schlatt2021kyc}. Since then, multiple other projects have combined anonymous credentials with blockchains~\citep{maram2021CanDID,halpin2020nym,abraham2020revocable,muth2022towards}. For instance, \citep{abraham2020revocable} use a blockchain network for managing timely attestations of non-revocation that the verifier can use even if it is off-chain, and \citep{muth2022towards} support the verification of \acp{vp} using \ac{cl}-based anonymous credentials in smart contracts.

\subsection{Limitations of deployed anonymous credential systems}
A core feature that is not yet accessible in the digital wallet projects we found supporting anonymous credentials is privately linking a software-bound credential (i.e., the holder's wallet app does not offer dedicated protection of binding secret keys but rather loosely stores them on the same layer as other application-level data, making it easy to extract and copy), such as a COVID-19 vaccination certificate, with a strongly bound (i.e., the holder's wallet app stores binding secret keys with additional protection against theft and copying, e.g., in an embedded secure element) government-issued digital~ID that relates to the same person. This would be an example of a cross-credential predicate, e.g., a comparison of the date of birth and name attributes on both credentials, without disclosing these attributes to the verifier. A national ID~card, for instance, often requires holder binding with a secret key stored in secure hardware~\citep{rosenberg2022zkcreds,bsi2019mobileidentity}. Including such features is generally considered desirable, particularly in regulated environments~\citep{rieger2021vaccination} and could be extended to combinations of even more attestations in a privacy-preserving way, such as in the verification of event tickets~\citep{feulner2022ticketing}. Moreover, binding only few ``core'' credentials to secure hardware while allowing other credentials to be linked to these strongly bound credentials offers apparent benefits: On the one hand, software-bound credentials can inherit hardware binding by being presented together with strongly bound credentials by proving the equality of a subset of identity attributes that uniquely identify an individual. On the other hand, when changing devices, it suffices to revoke and re-issue the strongly bound credentials, while the software-bound credentials can be copied into the wallet app on the new device.

Hardware security modules and embedded secure elements generally do not support \ac{cl} or \ac{bbs}+ signatures but only more common signature schemes such as \ac{ecdsa}. Private credential chaining, another example of a cross-credential predicate, is also not supported in Hyperledger AnonCreds but considered essential for large-scale adoption~\citep{aries2022chaining}. Moreover, the number of credentials that revocation registries for \acp{zkp} of (non-) set-membership, implemented via \ac{rsa} accumulators in Hyperledger AnonCreds~\citep{camenisch2009accumulator}, can manage is far too small to guarantee sufficient herd privacy: to allow the holder to prove that their credential is not revoked, he or she needs to store data that grows linearly with the number of credentials represented by the accumulator in their wallet app. More precisely, the holder needs to store a static~$2000$~bit integer for every credential represented through the accumulator. For a revocation registry that represents 10,000~credentials, the corresponding ``tails file'' hence already has a size of 2.6~MB~\citep{bcgov2021indyrevocation}.  Consequently, it is no surprise that the maximum size of the revocation registry is set to $2^{15}=32,768$ in \ac{aca-py}.

In practice, facing limited capacities of revocation registries with \ac{rsa} accumulators, revocation registries are split. However, this compromises privacy significantly: Consider an identification process that involves information from three different credentials, e.g., a national ID~card, a credit card, and a COVID-19 vaccination credential. Let $N$ be the size of the population that owns one of each of these credentials and $r$ be the maximum number of credentials that can be represented by a revocation registry. Then there will be $Nr^{-1}$ revocation registries for each of the attestation types, and approximately $k=N^3r^{-3}$ combinations of revocation registry~IDs that an individual can refer to when presenting the three attestations together. For instance, if $N=50$~million and $r=100,000$, then $k=125$~million, i.e., the combination of revocation registry~IDs is essentially a unique identifier appearing in every \ac{vp} in which a holder uses these three attestations and proves their non-revoked state because $k>N$. When $r=1$~million, we have $k=125,000$, so there will be around $Nk^{-1}=400$~people with the same combination of revocation registries, i.e., herd privacy guarantees are still relatively bad, particularly if additional credentials and, thus, further corresponding revocation registries are around. For $r=10$~million and, therefore, already close to $N$, we get good herd privacy since $k=125$ and there will be $400,000$ individuals with the same combination. Consequently, revocation registries should represent several millions of credentials rather than tens of thousands. To achieve this with the \ac{rsa} accumulator approach implemented in Hyperledger AnonCreds, a digital wallet would need to store more than one~GB of revocation-related data per credential, which can be considered impractical.

It also seems that the approach with special-purpose \acp{zkp} is difficult to adapt to post-quantum security: While \citet{dutto2020pqzkpvc} were able to reproduce the key properties of the above-mentioned anonymous credential schemes, such as selective disclosure, private holder binding, and private revocation, with plausibly post-quantum secure cryptography (lattices), signature and proof sizes are on the order of several hundreds of~MB. These figures are arguably not yet suitable for large-scale roll-out, particularly for digital wallets running on mobile phones.

\subsection{Academic proposals based on \acp{snark}}

The difficulty of extending approaches based on special-purpose \acp{zkp}, such as \acf{cl} and \ac{bbs}+-based anonymous credentials, to needs in large-scale adoption motivated research to construct anonymous credentials using generic \acp{zkp} like \acp{snark}. The following works have focused on \ac{snark}-based anonymous credentials systems:

\citet{kosba2015coco} were the first to propose the use of \acp{snark} for constructing anonymous credentials. They base the security of their approach, C$\emptyset$C$\emptyset$, on a universally composable (UC)~\citep{canetti2001uc} \acp{snark} construction that allows to prove generic statements in a modular way. The UC property is achieved through a lifting construction. Using \ac{snark}-friendly cryptographic primitives, C$\emptyset$C$\emptyset$ achieves proving keys of several hundreds of~MB and proving times on the order of tens to hundreds of seconds.

\citet{delignat2016cinderella} present a more practically-oriented approach towards data-minimal \acp{vp}, ``Cinderella'', as they turn widespread X.509~certificates into anonymous credentials using \acp{snark}. 
A ``prover can verify that he holds a valid certificate chain and a signature computed with the associated secret key, without actually sending them to the verifier''~\citep[p.~1]{delignat2016cinderella}. 
Cinderella implements selective disclosure, private credential chaining, and private holder binding to secure hardware on top of the existing X.509~certificate infrastructure. Their work can be considered a milestone for bridging anonymous credentials and legacy certificate systems for servers, yet does not bridge the gap to the multi-credential systems and general predicates envisioned in \ac{ssi} and desirable properties like accumulator-based revocation registries as it builds on \ac{ocsp}. Moreover, at the time of publication, proof generation took around 4.5~minutes on a quad-core Desktop~PC for a \ac{vp} involving a single X.509~certificate with an \ac{rsa} signature, and around 9~minutes for a chain of three certificates. Unfortunately, the code is not open source to the best of our knowledge.

\citet{schanzenbach2019zklaims} propose ZKlaims, a \ac{snark} based approach to anonymous credentials, for application specifically in the context of blockchain technology, where an efficient smart contract verifier needs to be implemented. Besides the efficient verification of \acp{vp} in smart contracts, their main focus is on the selective disclosure of attributes and the implementation of range proofs using \acp{snark}. 
ZKlaims does not consider several key components of \ac{ssi}, such as holder binding (particularly to secure hardware), revocation, as well as more advanced predicates and combinations of anonymous credentials for cross-credential predicates, for which private credential chaining is a special case.
\citet{li2020privacy} also discuss how privacy-oriented identity verification could look on blockchains using \acp{snark}. Similar to \citet{schanzenbach2019zklaims}'s work, a smart contract on a blockchain verifies \acp{zkp} about credentials. However, while the corresponding architecture with a smart contract verifier is discussed, there are no implementation details given; and there is also no connection to discussions in \ac{ssi} and the typical statements such as holder binding and non-revocation for which 
a \ac{vp} needs to provide evidence. \citet{yang2020zero} also describe an implementation based on \acp{snark} that attests identity claims, with an approach that stores blinded commitments to attributes in a smart contract managed by one or several issuers. This approach allows implementing revocation and makes attribute usage unlinkable. Yet, holder binding, credential chaining, or the scalability of revocation are again not discussed.

\citet{buchner2020zero} propose a more advanced approach toward \ac{snark}-based anonymous credentials. They mention key privacy features for these credentials, such as selective disclosure and private holder binding and also consider privacy-oriented revocation; yet in a setting where the revocation status of a credential is directly checked in an interaction between the verifier and issuer, which poses higher availability requirements on the issuer. While not detailing their implementation and leaving several design-related questions, the authors emphasize the potential advantages of transparent \acp{snark} compared to, for instance, Groth16 \acp{snark} in terms of coordination (trusted setup) and opportunities for post-quantum security. 
\citet{Deevashwer2022ZEBRA} also focus on blockchain-based applications of \acp{snark}-based anonymous credentials. They use \ac{snark} batch verification to reduce the costs of the on-chain verification of multiple \acp{vp}. Their implementation includes selective disclosure and private revocation. Hardware binding, credential chaining, as well as more general predicates and the corresponding tooling are also not discussed. 

\citet{rosenberg2022zkcreds}'s approach ``zk-creds'', besides the contribution by \citet{delignat2016cinderella}, arguably comes closest to ours. They demonstrate the practicality of a \ac{snark}-based approach to anonymous credentials that include many desirable features with a sub-second proving time on a laptop. Notably, they provide formal security proofs for their construction. Moreover, their implementation supports existing identity (ICAO passports). However, their focus is more on establishing the cryptographic foundations than on describing how to design required features and corresponding trade-offs in an \ac{ssi}-based approach. \citet{rosenberg2022zkcreds} implement private, non-interactive proofs of non-revocation using Merkle forests and allow predicate proofs that involve multiple credentials, therefore also facilitating private credential chaining. Many of the opportunities of their constructions are only briefly described in the paper, such as wallet-side scalability considerations of revocation, hardware binding, and a predicate proof of geo-location. We detail several of these discussions and also implement the geo-location proof in the form of a polygon inbound proof in Appendix~\ref{code:polygon}.
Similarly, \citet{yeoh2023fidoac} propose a framework that uses \acp{snark}-based anonymous credentials in an extension of the established FIDO2 protocol, adding active authentication to~\citep{rosenberg2022zkcreds}. While this construction resonates with hardware binding requirements and supports existing documents, it still leaves open how revocation, credential chaining, and custom predicates can be deployed at scale.

Recently, \ac{snark}-based constructions of anonymous credentials have also attracted industry: For instance, there are efforts to use zk-\acp{snark} for privacy-friendly authentication with Aadhaar credentials~\citep{mirror2023anonymousAdhaar}. Aadhaar refers to the Indian governmental digital identity platform that has faced several privacy incidents before~\citep{sedlmeir2021digitalidentities}. Moreover, there are several initiatives that use zk-\acp{snark}-based authentication in combination with blockchains, such as the Verida Wallet that integrates Polygon~ID~\citep{verida2023polygon}, or QuarkID that is building on the rollup zkSync. QuardID's wallet is allegedly planned for rollout in Buenos Aires~\citep{zkSync2023quarkID}.

Finally, we note that there are also hybrid approaches to constructing anonymous credentials. For instance, \citet{campanelli2019legosnark} propose LegoSNARK to implement more common and frequently needed privacy features, such as selective disclosure and private holder binding, with highly performant \ac{bbs}+ signatures. If more advanced predicates are required occasionally, the LegoSNARK approach allows revealing also blinded commitments to selected attributes in the \ac{vp}. These blinded commitments can then be used as inputs for a more flexible but also more computationally expensive general-purpose \ac{zkp}, such as a Groth16~\ac{snark}. For instance, there is an implementation of this approach by \citet{dockio2022bbslegosnark}. Similarly,
\citet{chase2016efficient} propose a hybrid approach, combining \ac{cl} signatures and algebraic circuits for more specialized proofs of knowledge, for instance, of an \ac{rsa} or \ac{ecdsa} signature, using 2-party computation with garbled circuits, which they argue has better performance than the \ac{snark}-based approach that alternative constructions like \citet{delignat2016cinderella,rosenberg2022zkcreds} pursue. Yet, they do not suggest a concrete implementation of anonymous credentials or evaluate the performance of their proposal and the size of the garbled circuits empirically.  
Similarly to \citet{camenisch2001efficient} and \citet{feulner2022ticketing}, the authors of this work also emphasize the significance of non-transferability of credentials, supporting our hypothesis that holder binding to secure hardware, e.g., a mobile phone's embedded secure element, is desirable in \ac{ssi} and particularly important for anonymous or data-minimized authentication.

%Besides the paradigm of anonymous credentials being issued by trusted institutions in a regulated environment and a focus on data-minimizing \acp{vp} towards (smart contract) verifiers, there is also a literature stream on how to make the issuance of anonymous credentials more decentralized and accountable with the help of blockchains~\citep{garman2013decentralized}. Recently, \citet{maram2021CanDID} proposed a decentralized identity system that can be used in blockchain applications. First, ``off-chain'' attestations and corresponding revocations can be verified by a blockchain through oracles that prove the validity of some \ac{tls}-based communication, with an \ac{ssl} certificate as cryptographic trust anchor. Sensitive oracle information is not disclosed; instead, one can use either direct remote attestation via \acp{tee}~\citep{zhang2016town} or via \acp{zkp}~\citep{zhang2020deco}. Similarly to \citet{delignat2016cinderella}'s approach, digitally signed information that exists on the web as of today could therefore be used in a way that retrofits the core capabilities of anonymous credentials. Moreover, \citet{maram2021CanDID} use \ac{mpc} to identify sanctioned identities based on ``off-chain'' identifiers even under small modifications (e.g., a spelling error in the name) and implement sophisticated key recovery mechanisms. 

\subsection{Research gap}
Our survey of related implementations and academic research demonstrates that there is not only an established academic discussion but also a high practical need for a flexible, extendable solution facilitating data-minimizing \acp{vp} in \ac{ssi}. Compared to these previous valuable contributions, we add details on implementation such as discussions of encoding aspects (Section~\ref{subsec:encoding}), the required capacity of revocation registries and how to improve revocation registries' practical capacity with Merkle tree-based accumulators (Section~\ref{subsec:revocation}), as well as a more detailed discussion of hardware binding and all-or-nothing non-transferability. We also provide more detailed performance analyses, incorporate an example of a complex predicate (the proof of geo-location) and discuss aspects such as issuer unlinkability and the role of the (trusted) setup that previous research has not yet covered. Lastly, we contribute novel insights into how \acp{snark} can facilitate designated verifier \acp{zkp} that reduce the risk of breaches of cryptographically verifiable data and the need for restrictive certification of relying parties in privacy-oriented \ac{ssi} solutions (Section~\ref{sec:discussion}). Our work hence focuses on merging the \ac{snark}-based approach to anonymous credentials with existing requirements learned from deployments of digital wallets in pilots, covering additional aspects of the \ac{vp} and opening the discussion on how to bring anonymous credentials to digital wallets at scale to researchers beyond pure cryptography.
\section{Design and Implementation}
\label{sec:design}

In the following, we describe \emph{Heimdall}, our implementation of anonymous credentials with \acp{snark}. The code implements a command line demo for different \acp{vp} that include revocation, credential chains, advanced predicates such as a proof of geo-location, as well as designated verifier presentations. It is accessible open-source at~\url{https://github.com/applied-crypto/heimdall}.

\subsection{Credential structure}
\label{subsec:structure}

Our credential design aims to be as simple and as general as possible. We opted for a binary Merkle tree-based approach for several reasons: First, it makes the construction that we present intuitive and demonstrates that the use of general-purpose \acp{snark} allows reducing complexity while still improving significantly the variety of features and several performance aspects compared to, for instance, anonymous credentials based on \ac{cl} and \ac{bbs}+ signatures.
Second, Merkle trees allows for a ``hybrid approach'' that offers selective disclosure capabilities even to holders with highly resource-constrained devices via Merkle proofs where \acp{snark} may not yet be practical to create. 
Third, the use of unbalanced trees can improve efficiency when presenting all meta-attributes, but only selected attributes of large credentials. Fourth, we can use the Merkle tree to structure the credential's metadata and attributes according to their meaning without the need to implement a complex (de-)serialization or lookups inside a \ac{zkp}, as involved in ASN.1 parsing in \citet{delignat2016cinderella} or JSON parsing for \acp{jwt}~\citep{zklogin,snark2021jwt}. In particular, we outsource the mapping of the different attributes to their semantic meaning to a ``schema'', with a hash of the schema or the corresponding \ac{url} referenced in the credential's metadata. We proceed similarly for other descriptions, such as a revocation registry. This approach is similar to the one implemented in \ac{aca-py}, where the credential schema and revocation registry are stored on a Hyperledger Indy blockchain~\citep{schlatt2021kyc,linux2022aries}.

The left half of the associated Merkle tree corresponds to metadata that will typically be verified in each \ac{vp}, including a unique revocation~ID (credential~identifier / serial number) a reference to a schema, a reference to a revocation registry, the public key for holder binding, and an expiration date. The right half of the Merkle tree represents the content, including all the attributes, e.g., for a national~ID. Note that while our implementation is based on a balanced tree, including eight slots for meta-attributes and eight slots for attributes, also unbalanced approaches are conceivable, for instance, if the number of attributes gets much larger. 

For our implementation, we used the \ac{zkp}-friendly Poseidon hashing algorithm. We chose these Poseidon hashing algorithm despite its relative novelty because it is used in several blockchain projects, such as the privacy-oriented Dusk Network~\citep{maharramov2019dusk}. If someone found a security issue with Poseidon, it would likely be used to exploit these projects that secure digital assets worth tens of millions of~USD and, therefore, arguably quickly discovered and addressed through a patch. Moreover Poseidon has recently attracted increased, unsuccessful attempts by cryptanalysts trying to break them~\citep{ashur2023poseidon,kovalchuk2021poseidon}.

\subsection{Encoding the attributes}
\label{subsec:encoding}

An essential part of defining a credential design that related work does not describe explicitly is to specify an \emph{encoding} for the different data types of meta-attributes and attributes. As we discussed before, Circom only supports integer values lower than a 254-bit prime number for Signals. Consequently, identity attributes need to be encoded using these integers. For dates and timestamps, for instance, the representation through a large integer that any general-purpose \ac{zkp} system needs to perform ``behind the curtains'' is relatively straightforward to implement, e.g., via a UNIX~timestamp. We decoded short Strings by representing every character $a_i$ by an 8-bit integer through UTF-8 encoding (128~possibilities) and then ``concatenating'' these via $\sum_i a_i\cdot 2^{8i}$ where $0\leq i\leq 31$ (for the case of our specific prime field; 31=$\tfrac{253}{8}$) also gives rise to a 1:1 mapping, such that these Strings can be represented directly as leaves of the credential's Merkle tree. This approach is valuable when certain predicates need to be computed from the corresponding \mbox{(meta-)} attribute in the \ac{vp}.
However, for Strings without initial length restrictions, we need to compress the potentially large raw attribute into a single integer with at most $253$ bits. When doing so in a collision-resistant way, the \ac{vp} can selectively disclose the leaf for the corresponding (meta-) data, proving that it is indeed part of the credential with the \ac{zkp}, and then attach the raw attribute to the \ac{vp}. The verifier can then apply the same encoding to the raw attribute to see whether the result is the corresponding leaf selected in the \ac{vp}. A straightforward way to obtain such a compressing and collision-resistant encoding is to use a cryptographic hash function that maps inputs of arbitrary length to $253$~bits. An alternative is reserving several leaves for a potentially large String and to split it into smaller Strings that can be encoded without compression via $253$-bit integers as indicated above, such that the attribute can then still be used for computing meaningful predicates.
It is important to note that the encoding logic in the verification process is happening completely outside the \ac{snark}. Consequently, also highly performant and established but less \ac{snark}-friendly cryptographic hash functions such as the \ac{sha256} could be used for this step. Finally, we used the encoding \texttt{True}~$\rightarrow1$ and \texttt{False}~$\rightarrow0$ to represent Boolean values, and multiplied Floats with a factor of $10^7$ before rounding.

\subsection{Defining schemas and revocation registries}
\label{subsec:schema}
As indicated in Section~\ref{subsec:structure}, a schema describes the semantic meaning of attributes and their position in the credential. In this sense, it serves as some kind of credential template~\citep{schlatt2021kyc}. For instance, a schema for a national~ID would describe the data types and positions associated with identity attributes. The schema would also describe the corresponding encodings individually if this is not defined on a higher level uniformly for each data type. As such, the content of the schema is relevant for the verifier, as it determines the positions of the identity attributes which the verifier asks in the proof request. The schema can be represented directly by a data format like JSON that allows for a collision-resistant serialization and, therefore, a hash-pointed link to a website or blockchain transaction that specifies the schema.
Similarly, the description of a revocation registry may include information about the issuer, policies underlying revocation, specific governance rules, update intervals, etc.

\subsection{Integrity verification}
\label{subsec:integrity}

The first (and probably most obvious) statement that the verifier expects to hold in a \ac{vp} is that the credential has not been tampered with. As in many other approaches to credentials, integrity (``tamper resistance'') is ensured through the issuer's digital signature. In general, creating a digital signature on a credential involves two key ingredients: (1)~a deterministic and collision-resistant serialization and compression of the credential into a short number (e.g., a $254$-bit number) and (2)~using a secret key to create the digital signature; often via some kind of exponentiation. The Merkle root of the credential (see Section~\ref{subsec:structure}) already provides the former. Verifying a single Poseidon hash involves 240~non-linear constraints in our implementation. Additionally, each step in the verification of a Merkle tree involves a \texttt{Selector} with 5~non-linear constraints to distinguish between the two cases for ordering, i.e., the relative positioning of the two values to be hashed. 
Furthermore, we selected the \ac{zkp}-friendly \ac{eddsa}-Poseidon digital signature mechanism, which is the reason why the issuer's public key consists of two 254-bit numbers, representing a point with two coordinates on the Baby~Jubjub elliptic curve. \ac{eddsa} signatures trace back to Schnorr's original work and are well established~\citep{bernstein2012eddsa,NIST2023dss}. The verification of an \ac{eddsa}-Poseidon signature in Circom involves 4,218~non-linear constraints. However, one of the most common digital signature mechanisms as of today is arguably \ac{ecdsa}. This signature mechanism leads to a substantially higher computational effort when generating the corresponding \ac{zkp} in a \ac{vp} (see Section~\ref{subsec:holder_binding} and Section~\ref{sec:evaluation}).

Owing to the fact that most of the metadata in a credential will be relevant in many \acp{vp} because of the verification of schema and revocation registry as well as the proof of holder binding, non-revocation, and non-expiration, we compute the full corresponding Merkle subtree for the metadata to validate its integrity in the \ac{snark}: This corresponds to $2\cdot2^2 - 1 = 7$ pairwise hashes when we have $8=2^3$ leaves representing metadata, as opposed to  $3\cdot n$ pairwise hashes when verifying $n$ individual Merkle proofs. 
By contrast, on the attribute side, we expect that we often only need to reveal a small number of attributes from the credential, so for optimizing performance, we only verify individual Merkle proofs for attributes that need to be revealed.

\subsection{Expiration checks}
\label{subsec:expiration}

Proving non-expiration without revealing the correlatable issuance or expiration date is arguably one of the easiest parts of implementing \ac{snark}-based credentials. As the Circomlib library~\citep{iden32022circomlib} already provides implementations of range proofs, the verifier can supply a timestamp of their choice as UNIX~timestamp in the proof request.
The holder then uses the timestamp as specified by the verifier as (private) input and also displays it as public output of the circuit. 
Within the circuit, the holder proves that the private expiration date as retrieved from the credential is indeed larger than the timestamp specified by the prover with Circomlib's \texttt{LargerThan} or \texttt{LessThan} component.

\subsection{Revocation checks}
\label{subsec:revocation}

We designed and implemented revocation as follows: As described in Section~\ref{subsec:structure}, each credential has a unique revocation~ID; for instance, the $i^\mathrm{th}$ credential created by the issuer could receive the revocation~ID~\mbox{$i-1$}. (Because we need to account for timing-related attacks, this is an over-simplification, and in practice the issuer would randomly select a revocation~ID in the range from $0$ to the maximum number of credentials represented by the revocation registry that has not been selected for a previously issued credential.) As proposed by the W3C~\citep{w3c2021revocation}, we use a binary sequence (BitString) to represent the revocation state for each credential. The $i^\mathrm{th}$ bit is $1$ if the credential with revocation~ID~i is revoked and $0$, else. We compress this binary sequence into a single hash value using a Merkle tree: Bits $0, \ldots, 252$ correspond to the first leaf, bits $253, \ldots, 505$ correspond to the second leaf, etc. A Merkle tree of depth $n$ can, therefore, represent $2^n\,\cdot\,253$ credentials. To check the revocation state in a non-private way in which the holder sends the credential to the verifier, a verifier would look at the credential's revocation~ID $i$ and inspect whether leaf number $i$~\textbackslash~$253$ (integer division) at position $i~\%~253$ (rest of the integer division, modulo operation) is $0$~or~$1$. Accordingly, using Circom components for verifying Merkle proofs, extracting a number's k$^\mathrm{th}$ bit (see Figure~\ref{code:extractKthBit}), and integer division with rest (see Figure~\ref{code:modulo}), the holder can prove that the credential is not revoked in a \ac{snark}. This is accomplished by utilizing that the revocation~ID that was already verified in the integrity verification part (Section~\ref{subsec:integrity}) and the Merkle proof associated with the corresponding leaf of the revocation Merkle tree as private input. Meanwhile, the public output only includes the current Merkle root of the revocation registry, thus ensuring a concise proof of non-revocation. The Merkle proof for a revocation registry that represents $253\cdot 2^n\approx 2^{n + 8}$ credentials involves verifying $n$ Poseidon hashes, whereas the verification of the integer division with rest and the extraction of the k$^\mathrm{th}$ add only a relatively small number of non-linear constraints (see Section~\ref{subsec:complexity})

\subsection{Holder binding (including secure elements)}
\label{subsec:holder_binding}

Implementing private holder binding is relatively simple with general-purpose \acp{zkp}~\citep{rosenberg2022zkcreds}. In essence, the holder proves that he or she is able to digitally sign a random challenge provided by the verifier. The public key for holder binding and the signed challenge are not communicated to the verifier; instead, the holder only proves that they could privately provide an input for the circuit such that the verification of a digital signature on the challenge with the binding public key as incorporated in the credential is valid. As for the issuer's signature (see Section~\ref{subsec:integrity}, we implemented the \ac{eddsa}-Poseidon digital signature scheme for efficient \ac{snark} generation.
Yet, as we detail above, storing secret keys in software is not sufficient for activities in strongly regulated areas. For instance, buying SIM~cards or opening bank accounts typically requires a ``high level of assurance'' in the European Union according to the \ac{eidas} regulation~\citep{schwalm2022eidas} that cannot be provided by keys stored in software, as they can be stolen or passed on relatively easily. Some security bodies also take the view that trusted execution environments like Android's Trusty do not provide sufficient security to achieve this level of assurance because several successful side-channel attacks and exploits have been detected~\citep{fischer2019new}. Only specific hardware, such as hardware security modules or (embedded) secure elements with highly restricted functionality are deemed sufficiently secure to provide such high levels of assurance~\citep[e.g.,~][]{bsi2019mobileidentity}. Yet, the secure elements that common devices like laptops or mobile phones carry today only support a very limited range of cryptographic operations. For storing a secret key and exposing the functionality of signing a challenge, this is the \ac{ecdsa} algorithm. \citet{0xPARC2022ECDSA} implemented \ac{ecdsa} verification with around 1.5~million non-linear constraints, as \ac{ecdsa} is a common signature mechanism on blockchains such as Ethereum. This makes the straightforward approach to \ac{snark}-based \ac{ecdsa} signature verification around~360~times more complex to prove than the \ac{eddsa} verification that we use in our sample implementation. Fortunately, there have been improvements that create some auxiliary private inputs to reduce the number of non-linear constraints by a factor of 10~\citep{github2022ecdsaoptimized} i.e., with this optimization it is only around 40~times more expensive to verify an \ac{ecdsa} signature than an \ac{eddsa}-Poseidon signature. However, there is some additional overhead for creating the auxiliary private inputs, which takes 2~seconds on the laptop used for the performance evaluations in Section~\ref{sec:evaluation}. 
%\jsnote{add raspi?}

It is important to note that by using an adequate governance approach, the verification of embedded secure elements' X.509~certificate chains
is not required to happen inside a \ac{snark} if unlinkability toward the issuer is not required. This is because the whole certificate chain for the secure element can be disclosed to the issuer, who publicly announces that it only binds credentials to key-pairs that are provably generated in secure elements from a trusted list of manufacturers as part of its governance policy.
In essence, \acp{snark} hence allow to draw a ``black box'' around the common challenge-response mechanism used for holder binding and avoiding replay attacks; with the opportunity to integrate any signature mechanism and in particular ones supported by the secure elements embedded in current generations of mobile phones. This is also the approach \citet{delignat2016cinderella} follow by compiling X.509 verification libraries into \ac{snark} provers and verifiers.
As such, \ac{snark}-based private holder binding can address a key shortcoming of approaches with special-purpose \acp{zkp} such as Hyperledger AnonCreds with \ac{cl} and \ac{bbs}+ signatures, as the corresponding signature mechanisms are not supported by practically deployed generations of secure elements and, thus, leave only the choice of either accepting a lower level of assurance or lowering the level of privacy by disclosing a unique identifier in the form of the public binding key (and potentially a device identifier included in the secure hardware's certificate chain).

Moreover, while \ac{bbs}+ signatures have been standardized by ISO and implemented on SIM~cards for testing purposes, and are performant enough to implement on embedded secure elements, we should foresee that addressing changing requirements through hardware with special-purpose \acp{zkp} is in general challenging in a landscape with rapidly changing technology and high inclusivity requirements: Even if microcontroller manufacturers would start integrating \ac{bbs}+ signatures in their secure elements today, it would take years until the number of end users with mobile phones without this functionality is negligible enough to ensure inclusivity. As we must also foresee new vulnerabilities found in cryptographic constructions or deprecations of established signature mechanisms owing to increasing computational power or emerging quantum computers, a modular anonymous credential construction that can more flexibly adapt to novel means of hardware binding may be desirable.

\subsection{Credential linking and credential chains}
\label{subsec:credential_chaining}

In real-world applications, another frequently required aspect is the combination of various attestations issued to the same individual or entity. For instance, when entering a facility that requires a proof of vaccination, a verifier may demand evidence that the digital vaccination passport (which is typically not strongly bound to the individual because it does, for instance, not include biometric information) refers to the same person that just demanded access~\citep{sedlmeir2021digitalidentities}. As other government-issued documents such as ID~cards often have higher binding strength (level of assurance), potentially also through hardware binding, it may make sense to prove that the first name, last name, and potentially the date of birth on the ID-card and the vaccination passport coincide, yet without leaking the sensitive (and irrelevant) name and date of birth directly to the verifier. Another frequent case of credential linking involves a proof that the public key for holder binding on one of the credentials is the same as the public key corresponding to the issuer's signature on the other credential. This is the building block of credential chains in which responsibilities are delegated from larger actors to smaller actors, e.g., from a government-controlled certificate authority to institutions on the national level to institutions on the local level to employees of these institutions, who then sign credentials on behalf of their institution and -- indirectly -- on behalf of the head institution on the national level. 
In this scenario, it may be prudent to conceal the issuer's public key for all credentials except the one at the top of the hierarchy. Without this precaution, it is possible that even if a \ac{vp} selectively reveals only an individual's date of birth and not their address from a national ID~card, the place of residence can still be deduced from the local authority that issued the national ID.

Credential linking can be achieved relatively simply by presenting both credentials individually and adding in both \acp{vp} a public output corresponding to the salted hash of attributes on both attestations (e.g., the first name). The circuit also needs to make sure that in both \acp{vp}, the same salt is used.
One can then either observe equality of the resulting hashes directly or proceed with a separate proof about properties of these salted hashes' pre-images, as would be the case with LegoSNARK~\citep{campanelli2019legosnark} or related approaches such as the one presented by~\citet{damgaard2021balancing}. We implemented the approach with direct equality checks on the salted hashes (outside the \ac{snark}) and used the signed challenge generated for proving holder binding as randomness/salt. For credential chaining, we conduct proofs of integrity, non-revocation, and non-expiration for the root credential and all intermediary credentials. The proof of holder binding is only conducted on the lowest level, as the binding keys on the other levels of the credential chain correspond to issuing keys and are, therefore, not shared with the holder.
Note that the holder needs to store all intermediary certificates, as well as corresponding revocation information, in their wallet to prove the validity of such a certificate chain. Consequently, the issuers' privacy on the lower levels may become an issue in some cases. We leave a discussion on how to improve the privacy of issuers for future work and only mention here that recursive \acp{snark} could provide an option at least in simple cases without revocation, with the holder receiving a \ac{snark} for authorized issuance instead of an intermediary certificate.

% \subsection{Remaining metadata}

% There are only few further checks that need to be conducted. One includes whether a credentials makes its holder eligible for issuing chained credentials, which we implemented by incorporating a binary value in the $6^{\mathrm{th}}$ leaf. Another is to selectively disclose the hash corresponding to the schema and revocation registry (or their \acp{url}) as described previously.
\section{Evaluation}
\label{sec:evaluation}

\subsection{Complexity of the statements to prove}
\label{subsec:complexity}

\renewcommand{\arraystretch}{1.2}
\begin{table*}[!hbt]
    \centering
    \resizebox{\linewidth}{!}{%
    \begin{tabular}{r|r||r|r|r|r|r|r|r|r|r|r|r|r|r|r}
        \multicolumn{2}{c||}{\textbf{Building blocks}} & \multicolumn{10}{c}{\textbf{Number of occurrences in corresponding scenario and contribution to number of constraints}} \\\hline\hline
        \textbf{Component} & \textbf{\# constraints} & \multicolumn{2}{c|}{\textbf{I}} & \multicolumn{2}{c|}{\textbf{II}} & \multicolumn{2}{c|}{\textbf{III}} & \multicolumn{2}{c|}{\textbf{IV}} & \multicolumn{2}{c|}{\textbf{V}} & \multicolumn{2}{c|}{\textbf{VI}} & \multicolumn{2}{c}{\textbf{VII}} \\\hline
        & & \# Occurrences & Constr. & \# Occ. & Constr. & \# Occ. & Constr. & \# Occurrences & Constr. & \# Occ. & Constr. & \# Occ. & Constr. & \# Occ. & Constr. \\\hline\hline
        Selector & 5 & 4 + 13 = 17 & 85 & 17 & 85 & 22 & 110 & 4 + 3 * 13 = 43 & 215 & 17 & 85 & 17 & 85 & 43 & 215 \\\hline
        %Num2Bits(3) & 3 & 1 & 3 & 1 & 3 & 1 & 3 & 3 & 9 & 1 & 3 & 3 & 9 \\ \hline
        Range proof & 252 & 1 & 252 & 1 & 252 & 1 & 252 & 3 & 756 & 1 & 252 & 1 & 252 & 3 & 756 \\\hline
        Division with rest & 252 & 1 & 252 & 1 & 252 & 1 & 252 & 3 & 756 & 1 & 252 & 1 & 252 & 3 & 756 \\\hline
        Poseidon hash & 240 & 1 + 4 + 7 + 13 = 25 & 6,000 & 28 & 6,720 & 30 & 7,200 & 4 + 3 * (1 + 7 + 13) + 2 * 2 = 71 & 17,040 & 25 & 6000 & 0 & 0 & 0 & 0 \\\hline
        extractKthBit & 1,012 & 1 & 1,012 & 1 & 1,012 & 1 & 1,012 & 3 & 3,036 & 1 & 1,012 & 1 & 1,012 & 3 & 3,036 \\\hline
        \ac{eddsa} signature & 4,218 & 1 + 1 = 2 & 8,436 & 2 & 8,436 & 2 & 8,436 & 1 + 3 * 1 = 4 & 16,872 & 1 & 4,218 & 0 & 0 & 0 & 0\\\hline
        \ac{sha256} hash & 29,636 & 0 & 0 & 0 & 0 & 0 & 0 & 0 & 0 & 0 & 0 & 25 & 740,900 & 71 & 2,104,156 \\\hline
        \ac{ecdsa} signature & 163,239$^\ast$ & 0 & 0 & 0 & 0 & 0 & 0 & 0 & 0 & 1 & 163,239 & 2 & 326,478 & 4 & 652,956 \\
        & (1,508,136) & & & & & & & & & & & & & \\\hline\hline
        \multicolumn{2}{r||}{\textbf{Total number of constraints}} & \multicolumn{2}{r|}{\textbf{16,037}} & \multicolumn{2}{r|}{\textbf{16,757}} & \multicolumn{2}{r|}{\textbf{17,262}} & \multicolumn{2}{r|}{\textbf{38,915}} & \multicolumn{2}{r|}{\textbf{175,058}} & \multicolumn{2}{r|}{\textbf{1,068,979}} & \multicolumn{2}{r}{\textbf{2,761,875}}
        \end{tabular}
    }
    \caption{Number of constraints for the most relevant basic building blocks and their occurrence in the basic scenarios. \qquad\quad $^\ast$ with preprocessing inputs.}
    \label{tab:componentsAndCosts}
\end{table*}

We specify the computational complexity for proof generation through the number of \ac{r1cs} constraints, which is approximately proportional to proving time (see also 
Figure~\ref{fig:performance}), decomposed by the different basic components.
Table~\ref{tab:componentsAndCosts} summarizes the number of constraints associated with each of the basic components and the total number for seven scenarios of a \ac{vp}, each of which involves the verification of integrity, non-expiration, non-revocation, and holder binding. For instance, a non-revocation proof with the Poseidon hash in a revocation registry representing $2^{13}\times252\approx2$~million credentials involves 13~Poseidon hashes and Selectors, an \texttt{IntegerDivisionWithRest} (see Appendix~\ref{code:modulo}), and an \texttt{extractKthBit} (see Appendix~\ref{code:extractKthBit}) component. There are also a few additional constraints associated with further operations, such as converting an attribute's position in the credential into a Merkle path via the \texttt{Num2Bits} component, which adds only $1$ constraint per hash in a Merkle proof. Note that combining several components in one circuit can even slightly decrease the total number of non-linear constraints that are a proxy for proving complexity and in particular proving time~\citep{albert2022distilling}, although the reduction is not significant.

The default setting (I) uses the Poseidon hash, the \ac{eddsa}-Poseidon signature, and a revocation registry that represents 2~million credentials. It discloses the value of a single attribute. (II)~corresponds to a presentation of all 8~attributes, (III)~to a revocation registry that represents more than 65~million credentials, and (IV)~to a presentation of three chained credentials. (V)~is the default scenario with \ac{eddsa}-Poseidon-based holder binding replaced by \ac{ecdsa}-based holder binding, (VI)~completely substitutes Poseidon and \ac{eddsa}-Poseidon by \ac{sha256} and \ac{ecdsa}, i.e. for the Merkle trees representing both the credential and the revocation registry, and (VII)~involves the presentation of three chained credentials of type (VI).

Similar to selectively presenting more than one attribute or increasing the number of credentials represented by a revocation registry, many further variations, such as increasing the number of leaves representing attributes from 8~to~32, only have a negligible impact on proving performance: This would merely add 245~non-linear constraints for each revealed attribute (one hash and one Selector), i.e., proving complexity is only increased by a few percent compared to the digital attestations with~8~leaves devoted for attributes. 

\subsection{Performance measurements}
\label{subsec:performance}

\begin{figure}[!tb]
    \centering
    \includegraphics[width=\linewidth]{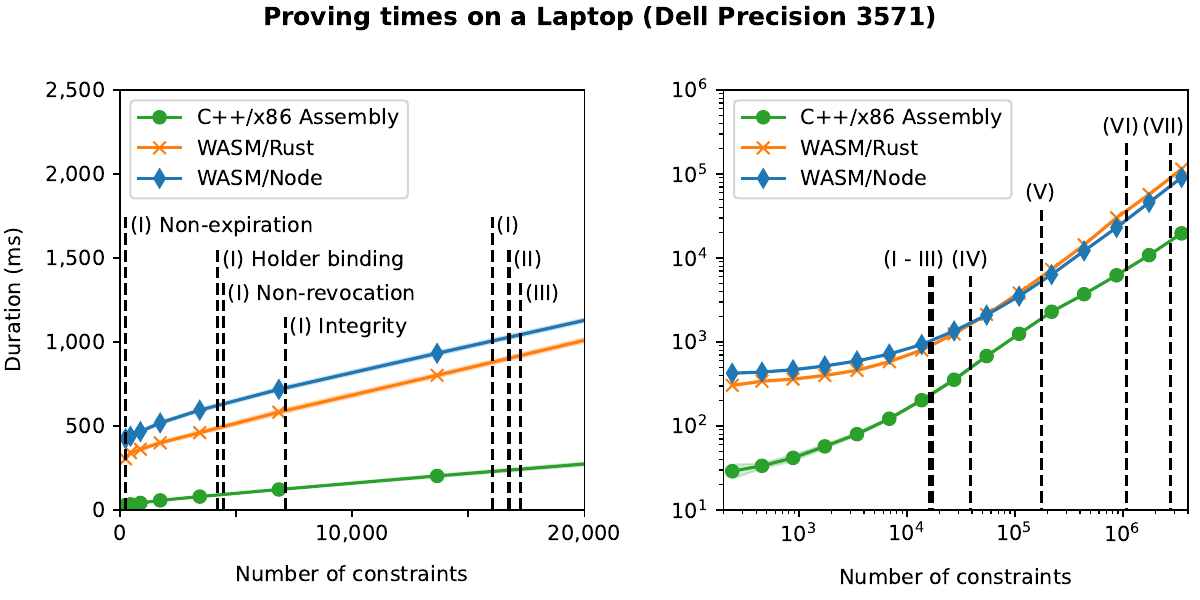}\\[0.2cm]
    \includegraphics[width=\linewidth]{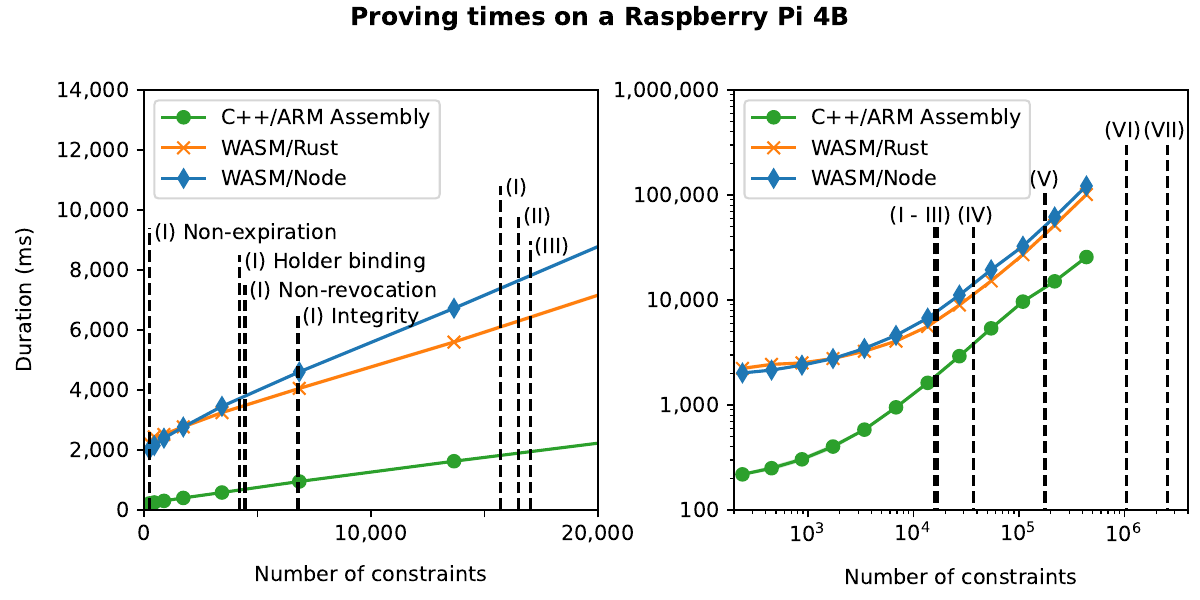}
    \caption{Proving time for different components of scenario (I) and the other scenarios on a virtual machine on a laptop and a Raspberry Pi~4B with different proof creation libraries.}
    \label{fig:performance}
\end{figure}

First, to obtain a lower bound for proving times, we tested the duration of proof generation on a high-end laptop (Dell~Precision~3571, Intel i9-12900H, 64\,GB~RAM, 2.5~GHz, 2.5\,GHz, 14~cores with a total of 20~threads on a Windows host, with 32\,GB~RAM, 7~cores, and 14~threads assigned to a Ubuntu 20.04~LTS virtual machine on which the tests were conducted) with different technology stacks (\Cpp/Intel~x86 Assembly, Node.js, and Rust, see Section~\ref{subsec:zkp}) and different complexities of statements by implementing circuits with a variable number of Poseidon hashes. The upper part of Figure~\ref{fig:performance} illustrates the corresponding results for a range between 240~and more than 3.5~million non-linear constraints. In scenario (I), i.e., when using a \ac{snark}-friendly hash function (Poseidon) and signature mechanism (\ac{eddsa}-Poseidon), proving time for a \ac{vp} that performs all metadata checks (integrity, non-expiration, non-revocation, holder binding) and reveals a single selected attribute is on the order of 300\,ms with \Cpp/Intel~x86 Assembly on the laptop, similar to \citet{rosenberg2022zkcreds}'s results. When using Node.js and Rust, proof generation duration is on the order of 1\,s and 2\,s, respectively. 
For the three chained credentials in scenario (V), total proof generation takes around 700~ms with \Cpp/Intel~x86 Assembly and 2\,s with Node.js and Rust on the laptop. For scenarios (VI) and (VII) that build on a more established hash function (\ac{sha256}) and signature mechanism that allows integrating the secure elements of existing hardware, proving times are considerably larger, yet still acceptable: Around 5\,s resp. 15\,s in \Cpp/Intel~x86~Assembly and around 30\,s  resp. 90\,s with Rust and Node. Proof sizes are on the order of a few hundred bytes, and verification takes around 1\,s with Node.js and 3\,ms in Rust, independent of the scenario.

\begin{figure}[!tb]
    \centering
    \includegraphics[width=\linewidth]{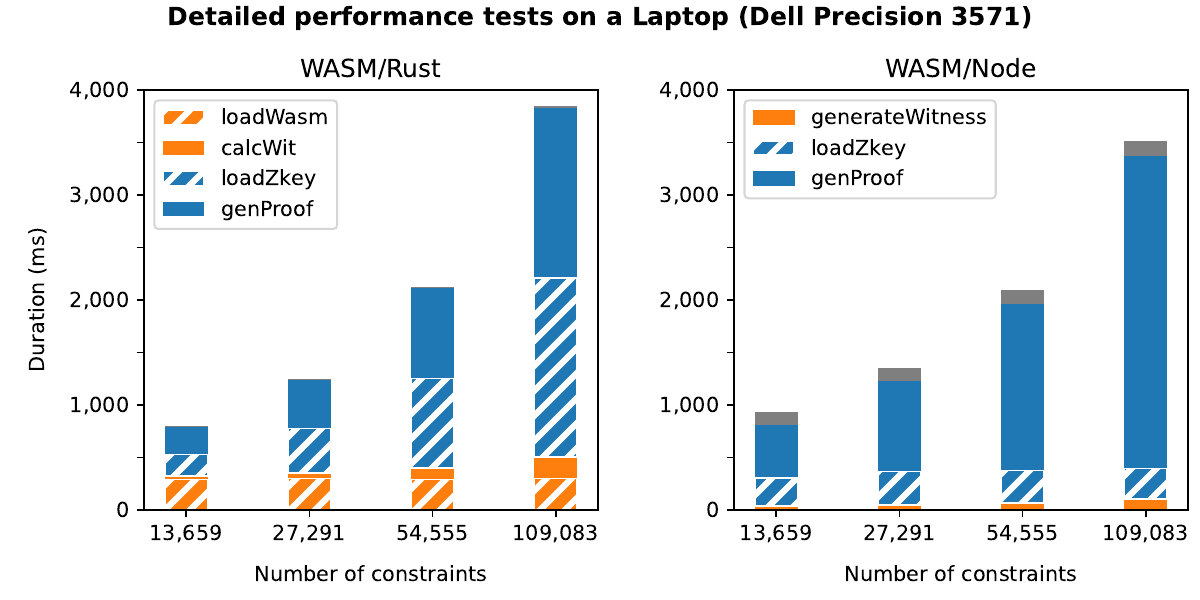}\\[0.2cm]
    \includegraphics[width=\linewidth]{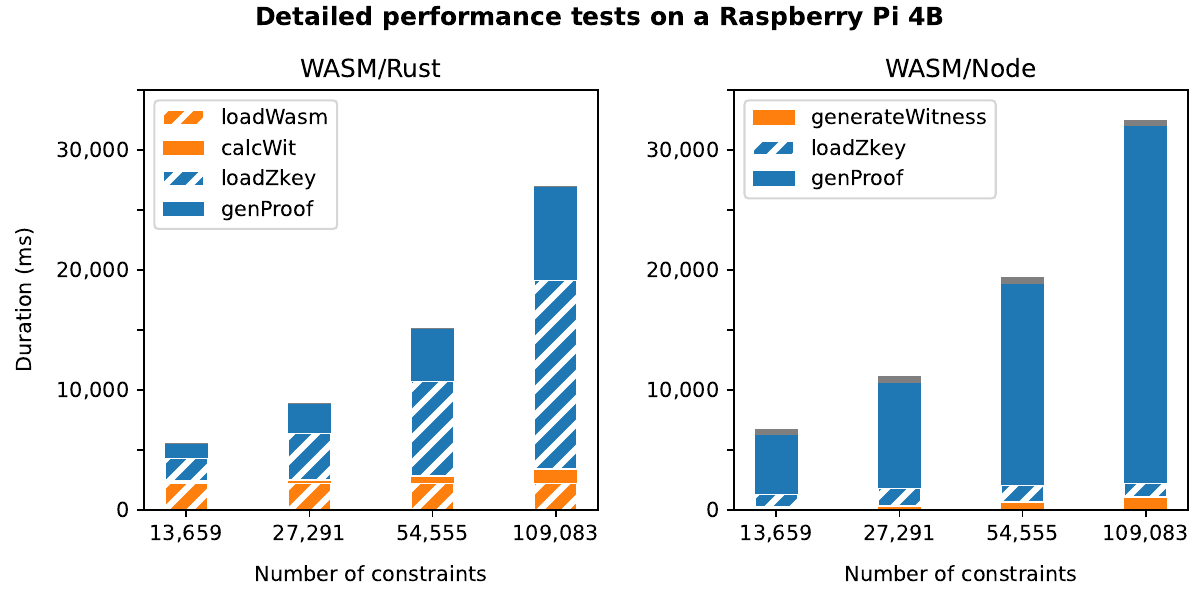}
    \caption{Duration of the different steps in the creation of the \ac{zkp} on a laptop and a Raspberry~Pi.}
    \label{fig:bars}
\end{figure}

As modern \acp{cpu} in mobile phones tend to have higher computational power than a Raspberry~Pi~4B~\citep{raspibenchmark}, we used a Raspberry~Pi~4B (Broadcom BCM2711, 4~GB~RAM, 4~cores with 1.5\,GHz and a total of 4~threads)  to obtain an upper bound on proving times for common edge devices. Similar to the laptop case, we performed performance tests for \Cpp/ARM~Assembly, Node.js, and Rust proving libraries (see Section~\ref{subsec:zkp}). We display the results in the lower part of Figure~\ref{fig:performance}. We found that proving time is around one order of magnitude higher than on the laptop, i.e., around 2\,s resp. 6\,s for scenarios~(I)~to (III) and 4~resp. 10\,s for scenario~(IV) for the \Cpp/ARM~Assembly and Node/Rust provers, respectively. A single proof of knowledge of an \ac{ecdsa} signature with~163k constraints takes around 30\,s with the Raspberry~Pi. Notably, as we illustrate in Figure~\ref{fig:bars}, when using Rust, a significant share of the total duration of proof generation is used for loading the \ac{wasm} file and the proving key both on the laptop and on the Raspberry~Pi. Read speed is naturally more limited for the Raspberry~Pi owing to the use of an SD~card as opposed to a \ac{ssd}. While total proof creation with the Raspberry~Pi in Node/Rust takes more than 5\,s even for the simplest \ac{vp}~(I), we see that the pure computation time for the proof in Rust (``genProof'') is only around 1\,s, with the major time spent on loading the \ac{wasm} code for witness generation (``loadWasm'') and loading the proving key for proof generation (``loadZkey'') from the file system. As modern smartphones tend to have considerably more computational power than a Raspberry~Pi, and significantly higher reading speeds as they use \acp{ssd} instead of a SD~Card, we hypothesize that proof generation time on a mobile phone when running the Rust~code natively may be considerably smaller. In contrast, for Node.js, by far the largest share is required for the computation of the proof from the witness. As such, it is not surprising that the \Cpp/ARM~Assembly prover can reduce proving time substantially. 
From these observations, we conclude that further optimizations are likely possible, such as re-using the witness generation and proving program once it is loaded into RAM, for instance, in a \ac{vp} that involves credential chains or multiple different credentials, may be able to further reduce the total time of cryptographic proof generation.

We have also started investigating performance on mobile devices. When running proof generation in the Browser or a react-native app on a mobile phone, proof generation is on the order of 7\,s for scenario (I) when using high-end mobile phones (Samsung Galaxy S10+ (8~cores, 1.9~--~2.7\,GHz, with a total of 8~threads) and iPhone 13) and between 15~and~30\,s for mid-range to low budget phones (Samsung Galaxy A6, Samsung~Galaxy~A32, Sony~Xperia~X~Compact), respectively. 
We noticed that both on the laptop and on Raspberry~Pi, proving performance in the Browser is worse than proving performance in Node/Rust, and that the choice of the Browser can make a considerable difference, with Firefox performing around 50\,\% slower than Chrome and Edge.
Using scenario (VI) and (VII) with \ac{sha256} and \ac{ecdsa} did not admit reasonable proof generation in a Browser; presumably for its significantly larger computation and memory requirements. Consequently, we are currently working on deploying proof generation on mobile phones with the Assembly provers available for Android and iOS as well as with Rust to explore proving times for common mobile phones in more detail. We are expecting proving times that are only a few times longer than proof generation on the laptop and considerably faster than on the Raspberry Pi. In this case, the verification of an \ac{ecdsa} signature with the reduced number of constraints should be substantially lower than 20\,s -- the duration with the \Cpp/ARM~Assembly prover on the Raspberry~Pi~4B.

Comparing our approach to the work by~\citet{Deevashwer2022ZEBRA}, which is also implemented with \acp{snark}-friendly primitives as far as the Ethereum virtual machine admits it, our simplest \ac{vp} has around 16,000 constraints as opposed to 62,000. Proving time for a single-threaded smartphone application with these 62,000 constraints is stated to be 6\,s, so according to~\citep{Deevashwer2022ZEBRA} we can expect around 1.5~seconds with a single-threaded prover for the simplest scenario~(I) and 4\,s for three chained credentials with \ac{snark}-friendly primitives.
In particular, \citet{Deevashwer2022ZEBRA}'s performance evaluation suggests that when implementing a multi-threaded \ac{snark} prover on a mobile phone with suitable software, proving time can likely be pushed below the 1~second range and therefore considered practical on a smartphone as of today, at least with \ac{snark}-friendly primitives. With this proving speed, an \ac{ecdsa} verification required for hardware binding would take less than 10\,s and can therefore also be considered practical on a mobile phone, considering the limited number of interactions requiring a high level of assurance users face in their daily lives.

Finally, owing to the advantages of universal \acp{snark} that we pointed out in Section~\ref{sec:background}, we tested proof generation with Plonk in Node.js and found that it performs around 50~times slower. Unfortunately, we did not find an optimized Assembly Plonk prover. Nonetheless, we encourage future experiments as there have been several improvements since, such as Turbo-Plonk and Ultra-Plonk~\citep{xiong2023verizexe}, with opportunities for optimizations via lookup tables that may substantially accelerate the hashing and signature verification components that are responsible for the overwhelming share of constraints, particularly for less \ac{snark}-friendly primitives.

\FloatBarrier
\section{Discussion}
\label{sec:discussion}

\subsection{Scalable revocation}
\label{subsec:revocation_discussion}

One of the core limitations of \ac{rsa} accumulator-based revocation when using \ac{cl} signature-based anonymous credentials is that every credential needs to be associated with a large integer to do a proof of non-revocation. The size of the integer ensures highly reliable privacy guarantees, so it cannot be simply reduced. Many alternatives~\citep{lapon2011revocationIdemix} or
optimizations of the underlying approach by~\citet{camenisch2009accumulator} have been suggested~\citep{hackmd2022zksam,nguyen2005accumulators}, and it seems that a combination of moving to pairing-based cryptography,  splitting revocation registries without compromising herd privacy, and distinguishing cases where no, few, and many credentials are revoked can indeed make revocation registries that cover several millions of credentials practical~\citep{hackmd2022nonrevoctoken}. Yet, this approach is arguably so complex that it has not been implemented in the larger \ac{ssi} projects thus far. Further, it seems that low client-side storage requirements can only be achieved by larger storage requirements on the accumulator side, which may complicate storage on a blockchain or other distributed, highly available systems.

In our implementation, issuers assigns every credential they issue a unique revocation~ID that is stored in one of the metadata fields. Our approach creates a proof of non-revocation via downloading the BitString or the full Merkle tree from the issuer's server (or a blockchain) and using it to create a proof that the bit at the specific leaf and position corresponding to the credential's private revocation ID is set to~1. One can readily see that the number of bits required to store the Merkle tree in uncompressed form is twice the number of bits of the leaves, i.e., around 4~million bits or 0.5~MB -- a reduction of a factor of 1,000 compared to the \ac{rsa} accumulator approach implemented in \ac{aca-py}. 
The \ac{snark}-based approach hence allows us to improve the information that needs to be stored on the holder side. On the other hand, considering the example of $65$~million credentials in a single revocation list, the $15$~MB for storing the full tree may still be considered too large to be practical. Fortunately, the Merkle-tree-based approach offers several further opportunities to improve specific resource tradeoffs: 
If storage and computation are expensive where the revocation registry is stored (e.g., on a permissionless blockchain), it suffices to only record the changes (``witness deltas'') to the revocation registry; for instance, an update transaction would record that $m$ credentials with IDs id$_0$, ..., id$_{m-1}$ have been revoked. The state would then include only $\log(N)$~bits per revoked credential, where $N$ is the total number of credentials represented by the revocation registry, and not require the computation and storage of hashes, reducing the amount of accumulator-related information to be stored significantly.

If, on the other hand, storage and computation on users' devices represent the bottleneck, there is also an opportunity to store the full Merkle tree corresponding to the revocation registry on some servers, and to query a Merkle proof for a certain leaf directly from the storage of the revocation registry. However, this could compromise herd privacy through correlating the query for a specific leaf to a \ac{vp}, such that a wallet would need to trust the corresponding service. Consequently, a more nuanced approach could be a hybrid form of the first and the second option: The wallet maintains a subtree of the Merkle tree (e.g., the left quarter when this includes the leaf in which the revocation bit for the credential under consideration is stored) locally by periodically pulling and applying witness deltas, and queries all Merkle nodes of the upper layers from a blockchain node or server. In the case of a revocation registry with $2$~million entries, this would mean that the wallet needs to store only around 125~kB and download 125~kB of data for updating the local revocation information that is necessary to produce a timely proof of non-revocation, without compromising on herd privacy guarantees toward the verifier and reducing herd privacy by a factor of~4 toward the server queried for the revocation registry.
    
If despite the availability of these tradeoffs, storage and computation of the revocation registry still are too resource intensive, the approach with general-purpose \acp{zkp} also allows for splitting the revocation registry into smaller parts without sacrificing herd privacy, similar to the approach in~\citet{hackmd2022nonrevoctoken}: The issuer could then provide separate Merkle roots, tag credentials and Merkle roots such that it is clear to which registry they belong, and digitally sign the revocation registries' Merkle roots together with a timestamp. Holders can then prove that their credential is non-revoked according to a Merkle root and a timestamp signed by the issuer and matching their ``tag'', without disclosing the Merkle root or the signature itself, thus preserving full herd privacy in the \ac{vp}. Note that this approach may increase proving time, as it involves an additional signature verification but reduces proving time only slightly (as proving time scales logarithmically with the number of credentials represented in the revocation registry). However, according to our performance evaluation in Section~\ref{sec:evaluation}, this additional signature check will take substantially less than a second even on a mobile phone, provided the signature mechanism is \ac{eddsa}-Poseidon.
On the other hand, splitting the revocation registries reduces the resources required for downloading and storing the current state of the revocation registry and re-computing the Merkle tree significantly (linearly). 

Another promising approach is based on the observation that many credentials, such as ID~cards or driver's licenses, are rarely revoked. For instance, there are close to 60~million driver's licenses in Germany, but only 0.2\,\% of driving licenses need to be revoked for a certain period of time (e.g., 3~months), with even fewer being permanently revoked. Consequently, storing only the revoked credentials' identifiers in the revocation registry may also reduce holder-sided storage requirements: As we need 26~bits to enumerate the credentials from 0~to 60~million, support for 50,000~revoked credentials at a time with an 8-ary Merkle tree created from these credentials' revocation IDs (8~times~26 bits fits into 253~bits, and $2^{16}$ is the first power of~2 larger than 100,000) implies a size of approximately 5\,MB. \footnote{Requires approx. $2^{16-3} \cdot 253 + 2^{16} \cdot \frac98 \cdot 253 \approx 10\cdot 2^{21} =1.25\cdot 2^{24}$ bits, i.e., 2.5\,MB.}
By comparison, the bit-string approach would have a size of approximately 16\,MB. ($2^{26}$ is the first power of~2 larger than 60~million)
\footnote{Requires approx. $2^{26}\cdot 2 = 2^{27}$ bits, i.e., 16\,MB.}. 
The corresponding \acp{snark} would then prove that the private revocation ID is in-between two adjacent revocation~IDs included in the revocation registry, with the depth of the corresponding Merkle proofs similar to the BitString example. However, it has been noted that non-membership proofs using sorted Merkle trees involve some inefficiencies, particularly when it comes to updates (complete re-computation) and proof size (overhead of 2x)~\citep{tomescu2022sortedMerkle}. Future research can investigate more efficient proofs of non-revocation facilitated by \acp{snark} and assess whether the additional overhead -- which can be mitigated by performing hourly or daily batch updates -- is acceptable and whether minimizing resource requirements on the issuer or holder side should be prioritized.

In general, our discussion of revocation mechanisms for \ac{snark}-based anonymous credentials suggests there are manifold ways to improve selected resource requirements. While it is unlikely that any of the above-mentioned approaches is Pareto-optimal, these examples illustrate that general-purpose \acp{zkp} facilitate a much richer design space for revocation that allows to flexibly address different resource restrictions.

\subsection{Designated verifier presentations}
\label{subsec:designated}

One issue that anonymous credentials cannot solve directly is that verifiers cannot be prevented from transmitting information (in particular, identity attributes) presented to them by holders to third parties. While impersonation attacks through replaying \acp{vp} can be avoided by using a random challenge in the proof request and demanding a proof of the capability to sign it with the binding secret key (see also Section~\ref{sec:background} and Section~\ref{subsec:holder_binding}), the verifier nonetheless can collect the revealed attributes and even use the \ac{zkp} attached to the corresponding \ac{vp} to provide evidence for the correctness of the data under consideration. Particularly in scenarios in which the corresponding attributes are highly sensitive, such as health-related personal information, plausible deniability or repudiability for the holder is desirable~\citep{hardman2020zkpsavvy,garrido2022polling}. The availability of machine-verifiable personal information is also one of the main reasons why the general idea of digital attestations in a digital wallet faces resistance from members of some net activist groups such as the German Chaos Computer Club~\citep{heise2022krawall}, and why the less flexible and convenient hardware-based solutions are sometimes considered a more privacy-friendly alternative because they do not transmit cryptographically verifiable attributes but instead only create a trustworthy communication channel.\footnote{Yet, facing the emergence of ``\ac{tls} oracles'' that \citet{maram2021CanDID} discuss in the context of digital identity infrastructures, this second argument can be contested.} 
%A verifier can prove to a third party that they received some data via a \ac{tls} connection with an entity with a given credential (e.g., a server with an SSL~certificate or a hardware-based~eID), in a bilateral, encrypted communication channel. More precisely, this approach utilizes that from the perspective of the verifier, the symmetric key used for encrypting the communication with the holder is derived from the holder's public key and the verifier's secret key., the verifier and the third party can engage in a multi-party computation to generate a shared secret that the verifier then uses for the handshake with the holder. Consequently, the verifier can reveal their communication protocol with the holder (or, using \acp{zkp}, a selected part of the communication) and therefore convince the third party about the correctness of these claims.

%One narrowly-scoped approach to provide plausible deniability despite the availability of cryptographically verifiable data is using differential privacy, i.e., adding noise to the attributes or predicates before revealing them. Indeed, general-purpose \acp{zkp} facilitate \emph{verifiable local differential privacy}, i.e., they can prove the correctness of noise generation and that the noise was indeed added to the correct value of the attribute~\citep{garrido2022polling}. However, in many scenarios, a trade-off in accuracy and, therefore, data quality will not be possible; particularly if a binary property or a sharp threshold (e.g., on age) is the basis for regulated authorization or process decisions.

A related, and arguably even more problematic, topic in the context of digital wallets is the tension field between users' wish to decide to who they want to disclose their information on the one hand, and security and privacy issues on the other hand~\citep{sedlmeir2022ssidps}.
In fact, one of the key challenges of adoption of self-managed identities involves controversies about how security risks arising from potential \ac{mitm} attacks should be balanced with end users’ informational self-determination as well as low entry barriers for relying parties~\citep{schellinger2022mythbusting}. A common solution is that holders must verify the identity -- and, therefore, trustworthiness -- of the verifier in a ``reversed \ac{vp}'' prior to the actual \ac{vp}. Omitting the identification of the verifier introduces significant security problems –- with a prominent example being the German ID-Wallet, which implemented anonymous credentials using Hyperledger Anoncreds~\citep{hyperledger2022anoncreds} based on \ac{cl} signatures. The rollout of the wallet was cancelled after security experts pointed out that it did not identify the verifier and was therefore vulnerable to \ac{mitm} attacks~\citep{schellinger2022mythbusting,lissi2022mitm}. For instance, an attacker could compromise a QR~code representing a link to the verifier's service that a holder scans to start an interaction with a verifier. Once the attacker notices activity, it interacts with a legitimate verifier to obtain their proof request (including the random challenge) and forward this proof request to the holder who followed the compromised link. As the holder believes that the attacker is the legitimate verifier, he or she creates a \ac{vp} for this proof request and sends it to the attacker. The attacker can forward the proof to the legitimate verifier. In other words, the attacker can use the \ac{vp} to impersonate the holder. This scenario gets even more concerning when the \ac{vp} is used to request a new credential from an issuer who first acts as a verifier to verify the eligibility for receiving a credential, as the attacker can make sure that this credential is issued to him-/herself, making impersonation possible for future interactions without the need for another \ac{mitm} attack and posing the risk of a gradual ``escalation of priviledges'' from accumulating credentials and using them for receiving new ones. The potential presence of \ac{mitm} attacks is therefore not only problematic for privacy reasons and in individual interactions but reduces the security and \emph{level of assurance} of identity documents in general. %\jsnote{similar regarding information disclosure towards issuers before issuance, but issuers will more likely be included in trust registries} 

Naturally, regulators demand reasonable protection against such \ac{mitm} attacks for scenarios or attestations -- at least for certain levels of assurance~\citep{eu2022guiding,verheul2021secdsa}. High bars on the certification of verifiers, however, inhibit the adoption and use of digital identities and users' control and informational self-determination. To give an example, the German~eID as implemented on a smart-card enforces that identity attributes can only be communicated to verifiers who have a certificate issued by a German national \ac{ca}~\citep{margraf2011new}. Getting these certificates is not only challenging because it requires the implementation and documentation of substantial security measures but also involves paying the certificate authority substantial amounts, with the outcome of the certification request being unclear. In the context of self-managed digital identities, this means that a digital wallet would not allow for sending a \acp{vp} to the verifier party unless the verifier can prove the possession of a corresponding certificate. This makes it difficult for self-managed digital identities to scale, for instance, to direct interactions between individuals or interactions involving smaller businesses and organizations as verifiers.

Projects that implement self-managed digital identities have hence suggested different ways to resolve this tension between adoption barriers owing to high certification requirements of verifiers on the one hand and security risks in the absence of verifier certification on the other hand. They are incorporating certification mechanisms that are relatively easy to access, such as using \ac{ssl} certificates for the identification of the verifier~\citep{bastian2022combination,lissi2022mitm}. There are also discussions that it may make sense to demand different levels of certification for different \acp{vp}; for instance, a \ac{vp} that only proves that a holder is older than 18~years may be relatively unproblematic even in the presence of a \ac{mitm} attack and, thus, require no or very little certification on the verifier’s side. Thereby, it is more accessible for scenarios such as buying alcohol at a bar or a small supermarket, where the verifier will unlikely have access to a sophisticated digital certificate. Yet, such a decision engine will arguably always trade security against low entry barriers. Determining the required certification level for the verifier based on the type and origin of revealed attributes and predicates also adds substantial complexity. Other approaches involve the holder more closely in the decision by asking them to accept certain risks (similar to circumventing an expired or non-existent \ac{ssl} certificate in the Browser). This approach seems challenging to implement, particularly, considering that there is a large global set of verifiers and corresponding processes in which authorizations could be accumulated with a snowballing-like approach.

A very elegant solution to this tension field are designated verifier \acp{vp}. The arguably simplest way to achieve this from the perspective of the holder in the context of X.509 certificates is to embed the targeted verifier's public key in their \ac{vp} in a tamper-proof way (e.g., in a message signed with the holder's binding secret key that includes their signed challenge and certificate) and to encrypt the message with the targeted verifier's public key. Thus, if a \ac{mitm} does not communicate its own public key but instead the legitimate verifier's public key, it cannot decrypt the \ac{vp} to extract and forward the relevant part. On the other hand, if the attacker communicates its own public key as the targeted verifier, such that it can decrypt and re-encrypt the holder's \ac{vp}, the attacker cannot modify the targeted verifier's public key in the \ac{vp}, such that the legitimate verifier will not accept the forwarded, re-encrypted \ac{vp}.
However, to the best of our knowledge, such designated verifier proofs have not been designed or implemented so far in the context of anonymous credentials. This is not surprising, as signing the message with the same key as the challenge for holder binding will again expose the holder's public key and, thus, a unique identifier. Moreover, while this design of a designated verifier \ac{vp} mitigates impersonation attacks from a \ac{mitm}, it still makes identity attributes in a \ac{vp} verifiable for any third party.

Outside anonymous credential systems, \emph{designated verifier} \acp{zkp} have been introduced before, describing \acp{zkp} that are only convincing for the intended recipient, but not for any third party that the \ac{vp} is potentially forwarded to~\citep{jakobsson1996designated,baum2022feta}. In fact, many forms of interactive \acp{zkp} are designated verifier \acp{zkp} because a third party to which the transcript of the interaction is forwarded cannot make sure that the transcript originates from unbiased randomness or that it is complete, i.e., that the responses that were not satisfying were not removed from the transcript~\citep{pass2003deniability}. However, non-interactive \acp{zkp} like \ac{cl}- and \ac{bbs}+ signatures and also \acp{snark} are designed to remove the inefficient, repeated interaction between prover and verifier, so they do not have this property by design. For \acp{snark}, the \ac{crs} reflects the fact that the \ac{zkp} is convincing toward any verifier~\citep{canetti2007universally}. Fortunately, with a simple trick, the designated verifier property can be added to a \acp{snark}-based \ac{vp} by proving the following statement~\citep{buterin2022designatedverifier}: \emph{Either I possess credentials that satisfy all the requirements of the proof request, including the correctness of the revealed attributes, or I know the verifier’s secret key}. As a common security requirement of \acp{snark} is that they are non-malleable (i.e., one cannot efficiently create a proof for certain public outputs from any number of other proofs with other public outputs), this approach also solves the challenge of creating a tamper-proof envelope without disclosing a public key linked to one of the credentials used in the \ac{vp}. However, the proof still needs to be encrypted with the designated verifier's public key.

The verifier's secret key can correspond to either a public key in a relevant, long-term credential or an ephemeral keypair used for only one specific interaction. If the holder sends the designated verifier \ac{snark} to the designated verifier, this verifier will be convinced because they know that they protected their own key-pair, i.e., the holder cannot have access to the corresponding secret key and, therefore, must have access to credentials satisfying all requirements. In contrast, any other party knows that the designated verifier \ac{snark} can be created trivially by someone who knows the corresponding secret key. In the case of an ephemeral designated verifier key-pair, any third party will therefore not be convinced. While the designated verifier proof may be more convincing for third parties when the designated verifier key-pair is long-lived and related to a reputed entity, such entities reputation would quickly disappear once they get known for forwarding designated verifier \ac{vp}.

% One straightforward way to implement designated verifier \acp{vp} in the \ac{snark}-based approach is as follows: Let $a_1,\ldots,a_n$ be all the assertions that the \ac{vp} needs to satisfy (e.g., integrity, non-expiration, non-revocation, holder binding, etc.), and let us assume for simplicity that $a_i === 1$ for all $i\in\{1,\ldots,n\}$ was asserted in the classic \ac{vp}. By sequentially multiplying the $a_i$ (we can only do one multiplication a step because of the underlying \ac{r1cs} constraint system), the correctness of the \ac{vp} can be asserted simply by demanding \mbox{\texttt{a === 1}}, where $a=\prod_{i=1}^n a_i$. \mbnote{Ich würde diesen Abschnitt glaube ich komplett weglassen. Ggf. sogar den kompletten Absatz}
% Let $b$ the result of a component that verifies whether an input to the \ac{zkp} was the digital signature of the challenge specified in the proof request with the verifier’s secret key, where \texttt{b = 1} represents a valid signature and \texttt{b = 0} an invalid signature. Instead of demanding \mbox{\texttt{a === 1}} in the ``classical \ac{vp}'', we then simply assert that \mbox{\texttt{a + b - a * b === 1}}, which is the arithmetization of demanding \texttt{a == 1 OR b == 1}, i.e., either the holder knows a valid \ac{vp}, or it can generate a signature that only the verifier is supposed to be able to generate. 

We implemented designated verifier \acp{vp} in Heimdall in the following way: We first extended the (private) inputs to a \ac{vp} by an ``enabled'' bit and a digital signature on the challenge with a secret key that is supposed to be created with the designated verifier's secret key. The public outputs are extended by the designated verifier's public key. We then mediated the signature verification of the credential with the signature verification of the designated verifier through the built-in ``enabled'' bit of the \ac{eddsa} verification circuit: Only one of them needs to be correct to generate a valid \ac{snark}, i.e., a valid designated verifier \ac{vp}.
The prover can then decide to either enable the verification of the issuer's digital signature on the credential or to enable the verification of the issuer's signature on the challenge with the (designated) verifier's public key. Any holder that wants to convince the verifier will choose to enable the verification of the issuer's signature, as they cannot create a signature on the challenge with the verifier's public key. On the other hand, a verifier who is not the designated verifier may suspect that the holder chose to verify the digital signature with the specified (other) designated verifier's secret key. Therefore, he or she did not verify the issuer's digital signature on the credential. The verifier knows that the holder can create an arbitrary credential that attests any claim he or she likes, such that the verifier will not find any assertion of the \ac{vp} trustworthy.
With this simple implementation, the designated verifier presentation for a single credential only involves 4,219 additional non-linear constraints: 4,218~constraints for the additional \ac{eddsa} signature verification and one constraint to check that \mbox{\texttt{enabled * (enabled - 1) === 0}}, i.e., the \texttt{enabled} input is indeed a bit.

\subsection{Privacy with respect to the issuer}

Some existing large-scale implementations of anonymous credentials use a so-called \emph{link secret} for holder binding~\citep{schlatt2021kyc,zundel2021transferability}. In essence, this works similarly to the private holder binding that we described in Section~\ref{sec:design}, with the main difference that the same binding link secret can be used in many credentials, yet every issuer includes it only in blinded form (more precisely, a salted hash with credential-specific salt) in the credential. This allows the holder to avoid being correlatable not only by verifiers (through the private holder binding that we described in Section~\ref{sec:design}) but also by issuers, which seems relevant when aiming for an increasing number of credentials and issuing parties in a digital identity ecosystem. Yet, the desired ``all-or-nothing non-transferability''~\citep{camenisch2001efficient,feulner2022ticketing} that binding all credentials to the same link secret or key-pair should enable is not appropriately met with this approach as deployed in, for instance, the Hyperledger AnonCreds~\citep{hyperledger2022anoncreds}. The reason is that these implementations, the holder does not prove to the issuer that the link secret to be incorporated in blinded form in the credential is the same as the link secret in the holder's other credential(s). A malicious holder can therefore make the issuer include another holder's link secret in their credential, rendering a core aspect of the link secret ineffective.
With \ac{snark} and only a very small adaptation of the ``standard'' features of our approach to anonymous credentials that we described in Section~\ref{sec:design}, it is very easy to force a holder to use the same link secret for each of their credentials if desired. We just need to add a particular \ac{vp} to the issuance process: First, in a \ac{vp} towards the prospective issuer, the holder additionally outputs a salted hash of the public binding key associated with of the holder's existing credentials, with the salt including randomness from both the holder and the verifier (see, e.g., Section~\ref{subsec:credential_chaining}). The holder can then send the issuer the root hash of some small sub-tree of the meta-data of the credential that they would like to have issued (e.g., including the binding public key and the expiration timestamp, with enough precision to have sufficient entropy, or using the one meta-data leaf that is currently still empty to include some randomness) and a \ac{zkp} that the public binding key included in both hashes is the same. 
The issuer can then include this sub-root in the credential it signs, therefore provably binding the credential to the same key-pair as the credential that was previously presented, without learning the corresponding binding key (or link secret).
With similar ideas, the issuer can generally include selected attributes from a holder's other credentials without learning what they are. This approach may also be valuable when the binding key-pair needs to be bound to secure hardware: An issuer who wants to issue a credential that can satisfy a high level of assurance needs to make sure that the secret key associated with the binding public key it signs was indeed generated in trusted, certified hardware. To date, this is only possible by verifying the corresponding hardware's attestation chain, i.e., a chain of X.509 certificates. Given the opportunity to ``privately'' transfer this binding key to other credentials, a single issuer could verify this attestation chain (without \acp{zkp} and ``transform'' it into a credential, such that all other issuers can then rely on binding their credential to the same hardware without the holder disclosing the corresponding public binding key again.

\subsection{Arbitrary predicates}

\subsubsection{Polygon inbound/outbound proof}
\label{subsubsec:polygon}
With a standard \ac{vp} doing all the checks that a verifier can reasonably expect (integrity, non-expiration, non-revocation, holder binding), all the meta-attributes and attributes are available as private inputs, i.e., parameters, for further predicates.  
As an example that illustrates the generality of predicates that one can easily implement with general-purpose \acp{zkp}, we implemented a proof of geo-location in the form of a polygon inbound proof: Given two coordinates~$x$,~$y$ in the Euclidean plane (or, in approximation, on a small area on earth that can be considered flat), one can prove that for a given polygon as specified by the verifier in the proof request, $(x, y)$ is inside (or outside) the polygon.
Given some efficient C~code that determines whether a point is inside or outside a given polygon~\citep{franklin2006pnpoly}, the implementation in Circom is straightforward (see Appendix~\ref{code:polygon}). Every vertex of the polygon contributes 333~constraints (mainly responsible are the 64~constraints for each of the 5~comparators). Consequently, an inbound/outbound proof for a given polygon with~50~vertices adds only 16,650~non-linear constraints when using comparators for 64~bits, which arguably allows for a sufficient degree of precision. According to our performance evaluation in Section~\ref{sec:evaluation}, this would increase proving time by less than a second on a high-end mobile phone.

\begin{figure}[!tb]
    \centering
    \includegraphics[width=0.7\linewidth, trim=2cm 3cm 2cm 2.5cm, clip, page=3]{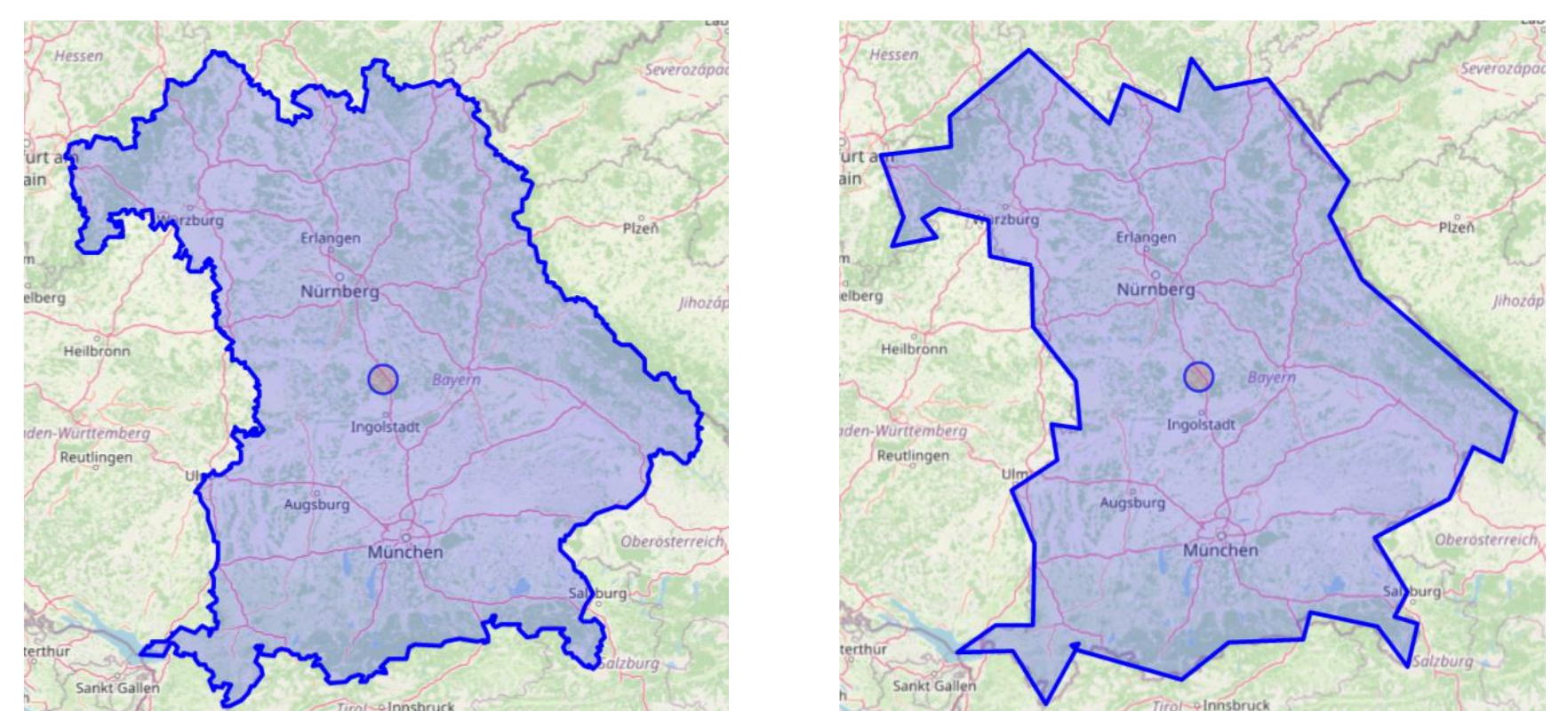}
    \caption{Approximating the boundary of Bavaria (left-hand side) through a polygon in the Euclidean plane with 50~vertices (right-hand side) such that a location in Bavaria can be efficiently verified with the polygon inbound proof (``leather pants predicate'').}
    \label{fig:leatherpantspredicate}
\end{figure}

There are several conceivable practical cases where an implementation of the polygon inbound/outbound proof can be useful. 
For demonstration, we included the coordinates of an individual's place of living in the credential (e.g., a national~ID) and used them to prove that the individual lives in Bavaria as an example of a certain city or federal state (see Figure~\ref{fig:leatherpantspredicate}). 
This predicate can be used to prove an authorization to claim certain benefits. Another example are regional energy markets where -- based on its location -- an intermittent source of green electricity like a roof solar plant can register to offer its electricity or flexibility~\citep[e.g.,~][]{antal2021blockchainlocalflex, mengelkamp2018designing}. Some research considers these local markets, on which energy assets can register autonomously, as a promising way to improve the share of renewables in the grid. Often, these markets are blockchain-based~\citep{strueker2019blockchain}, which means that the disclosure of sensitive information during the registration process is particularly problematic; even more for small assets owned by individuals. As we use Circom for the implementation of our prototype, we were also able to create a corresponding \ac{snark} verifier smart contract for Ethereum fully automatically, such that our \ac{snark}-based \acp{vp} can be verified by a smart contract -- provided the smart contract provides a source of pseudo-random challenges, e.g., using the hash of the current block, and controls for double-use to mitigate replay attacks.

Our prototype implementation also provides a generic, customizable circuit that verifies the integrity of all metadata and content data as well as non-expiration, non-revocation, and holder binding. When all content data has been verified in the circuit, a novel predicate can be implemented very easily, as only the predicate with private inputs \texttt{attribute[0]} to \texttt{attribute[7]} needs to be implemented. Consequently, for a customized predicate that adds up the first, second, and fourth attribute, the only thing the verifier would need to implement is a new output signal \texttt{returnValue} and assign \texttt{returnValue <== attribute[0] + attribute[1] + attribute[3]}.

\begin{comment}
[[10.4544399,47.5557964],[11.2698847,47.3975653],[11.6361799,47.5945549],[12.2039614,47.6067646],[12.2570286,47.7430345],[12.7811652,47.6738182],[13.0066207,47.4643842],[13.0807484,47.6870338],[12.9052586,47.7234349],[13.0033609,47.8500223],[12.7581257,48.1260686],[13.329798,48.3235141],[13.5089626,48.5905995],[13.730512,48.5147674],[13.8395518,48.771618],[12.6555517,49.4347994],[12.4005551,49.7538049],[12.5476758,49.920496],[11.9349229,50.4236526],[11.5194652,50.3739938],[11.3467213,50.5214416],[11.2531027,50.2678525],[10.8304723,50.3927122],[10.7174795,50.2043467],[10.1062239,50.5632937],[9.5013397,50.2431399],[9.5130437,50.0943483],[8.9763497,50.0497851],[9.1505794,49.7427032],[9.0664847,49.6022936],[9.4061438,49.6454555],[9.2956877,49.7404603],[9.4224965,49.789454],[9.7999073,49.730973],[9.8117746,49.5556982],[10.1212713,49.5110265],[10.1248147,49.1988368],[10.4568403,48.9204496],[10.4951298,48.6871989],[10.2687047,48.7035845],[10.3120672,48.522946],[9.9674293,48.3742096],[10.1407065,48.0977316],[10.1301574,47.6768474],[9.840425,47.677494],[9.550566,47.5371757],[9.9704798,47.5458589],[10.2323482,47.2705791],[10.4544399,47.5557964]]
\end{comment}

\subsubsection{Modularity and private credential bundling}

Selective disclosure and predicates can easily be extended to involving meta-attributes and attributes from multiple credentials. One approach to minimizing the \emph{verification} effort would be implementing a circuit that takes multiple credentials, the corresponding revocation lemmas, and the verifier's challenge as input and that outputs a single \ac{snark} that attests the validity of all input credentials and the result of the inter-credential predicate. In this case, the verifier would only need to verify a single proof. Yet, this approach seems to be less modular than presenting each credential individually and revealing salted hashes of the attributes that are needed for computing the predicate. In a second step, the holder can then prove the correct computation of the predicate based on pre-image proofs for the salted hashes that were previously revealed. Note that this approach is similar to the one taken by LegoSNARK~\citep{campanelli2019legosnark}, with the exception that in LegoSNARK, the initial \ac{vp} that outputs blinded commitments to attributes is not based on general-purpose \acp{zkp} but on \ac{cl} or \ac{bbs}+ anonymous credentials and blinded commitments to attributes. Only the second step that creates pre-image proofs for the commitments to use identity attributes in predicates is \ac{snark}-based to have the opportunity to compute arbitrary predicates across different credentials. An even higher degree of modularity at the cost of a larger number of proofs to be verified could be achieved by decomposing the verification of a credential (e.g., into individual circuits for integrity verification, holder binding, etc.) which are then linked via a salted hash in the same way we link credential chains.

\subsection{Trusted setup and flexibility of verifiable presentations}

So far, our performance analyses were mostly considering Groth16 \acp{snark}. One significant shortcoming of this type of \acp{snark} is the circuit-specific trusted setup, i.e., different types of \acp{vp} require different proving keys that need to be generated in an \ac{mpc} and transferred to the holder's wallet before generating a \ac{vp}~\citep{groth2016size}. For the initial bootstrapping of a system of digital wallets that can create \ac{snark}-based \acp{vp}, a few pre-defined proving keys, associated with common presentation types (e.g., a single attribute or two attributes revealed, where non-expiration, non-revocation and hardware binding, is always verified), hard-coded into the wallet or available as plug-in, may be sufficient. Yet, with a growing number of different credentials, a growing diversity of presentations (flexible number of attributes revealed), and verifier-specific predicates emerging, this solution is arguably not adequate anymore. One option could be to distribute the proving key with the proof request during the \ac{vp}. In a local area network, the large size of the proving key may not be problematic when a local bilateral connection is set up. For wide area network, Bluetooth, and \ac{nfc}-based transfer, several~MB for a proving key (and even hundreds of MB to few~GB when using \ac{sha256} for hashing and \ac{ecdsa} for digital signatures) are arguably not practical. Moreover, in this case, a sophisticated mechanism that gives users an opportunity to verify that no more information is requested than the proof request displays on their screen is required. The certification of proving keys that a wallet accepts, including a corresponding human-readable description, seems an obvious approach here; yet certification may make generic predicates less accessible to verifiers. What seems a more attractive option here that also provides plenty of avenues for future research is that after the relatively general verification of the metadata, the selected attributes and derived predicates could be described through a standardized format. This format allows for symbolic operations on the credential attributes, such that this description of the proof request allows for an automatic derivation of (1)~the corresponding circuit (constraint system) and (2)~the corresponding prompt that asks users for their consent. However, with this dynamic approach and a correspondingly large number of \acp{vp} types, a circuit-specific trusted setup that requires the transfer of the corresponding proving key seems impractical. On the other hand, general-purpose \acp{snark} frameworks with universal trusted setup, like Plonk and its successors, or with transparent trusted setup still involve a computationally and memory-intensive preprocessing step that is impractical to conduct on users' mobile phones. Arguably, zero-knowledge virtual machines, which conduct the setup process for the different instructions of a CPU to facilitate the dynamic verifiable execution of any input program, represent the most elegant approach. However, this approach seems to involve even higher proving times in both academic research~\citep{heath20202bubbleram} and industry deployments~\citep{risc0zkVM}.

\subsection{Programmable accountability}

Related research on anonymous credentials has sometimes discussed additional means to increase accountability, e.g., through anonymity revocation~\citep{camenisch2001efficient} and advanced features such as public auditability~\citep{brorsson2023papr}. While \ac{cl}-signatures can already support some type of anonymity revocation, in general, different cryptographic building blocks, such as verifiable secret sharing, have been used to increase the degree of different stakeholders' accountability. Supposedly, many of these means and in particluar the latter example can be implemented flexibly with \acp{snark} by building on verifiable encryption, which has been implemented in Circom in the form of ElGamal encryption~\citep{circom2020elgamal}. Verifiable encryption and/or the verifiable secret sharing of a subset of identity attributes included in a credential may also be helpful to construct means of \emph{identity escrow} in which a verifier can be sure that in the case of an incident, they could approach some institutions to obtain additional information about the identity subject that make it uniquely identifiable.
\section{Conclusion and Avenues for Future Research}
\label{sec:conclusion}

This paper has highlighted several key areas where general-purpose \acp{zkp} can address the shortcomings of existing implementations of privacy-oriented digital identity infrastructures. In particular, they can help address key requirements of anonymous credentials in self-managed digital identity projects that previously have been pointed out by various sources~\citep[e.g.,~][]{feulner2022ticketing, schlatt2021kyc,sedlmeir2022ssidps,schellinger2022mythbusting, hardman2020zkpsavvy} but that are not present even in advanced solutions like such as \ac{aca-py}~\citep{young2022real}. We illustrated that the key features that anonymous credentials need to support broad adoption in practice can be implemented with relatively limited effort using \acp{snark}. Related research such as work by \citet{rosenberg2022zkcreds} has already provided provable security for similar approaches, and we argue that the universality of the existing tooling makes implementations more flexible, audits easier to perform, and core cryptographic components more battle-tested with general-purpose \acp{zkp} than with approaches such as \ac{cl} and \ac{bbs}+ signatures.

The main limitation of \acp{snark} is arguably the computational complexity of proof generation~\citep{thaler2022snarkperformance}. We conducted several performance tests that suggest that even with our illustrative design that did not focus on performance, the general-purpose \ac{snark} tooling implemented by industry makes the speed of proof generation for a data-minimizing \acp{vp} practical on mobile phones as of today when using \ac{snark}-friendly cryptographic primitives. Moreover, we illustrated how general-purpose \acp{zkp} do not only improve on aspects such as private scalable revocation, hardware binding, and credential chaining or much more general predicates but also bring unprecedented opportunities such as plausible deniability and expanding the solution space for anonymous credentials in the tension field between user control and low entry barriers on the one side and the risk of \ac{mitm} attacks on the other side. Finally, we pointed out that there are still many open questions that provide promising avenues for future research, such as facilitating practical performance also for data-minimal \acp{vp} in settings where \ac{snark}-friendly hashing algorithms and digital signatures are not accessible or when using transparent \acp{snark} on mobile phones. Indeed, recent performance analyses suggest that transparent SNARKs outperform trusted setup SNARKs like Groth16 by more than an order of magnitude in proving time~\citep{ethresearch2023snarkbenchmark}, and while their proof size of tens of~kB may be challenging for many blockchain applications, it seems well-suitable for the bilateral interactions that are typical of \ac{ssi}-based digital identity management. However, to date, these projects often choose different security parameters~\citep{thaler2022snarkperformance} and are also not always zero-knowledge, although ``upgrading'' them to zero knowledge is relatively straightforward~\citep{bensasson2018scalablepreprint}. Future research that looks into alternatives with lower prover overhead may also consider general-purpose \acp{zkp} which are not succinct: While \ac{zkp} verification time should arguably not be superlinear in the complexity of the verification program, sublinearity also does not seem necessary for \acp{vp}. As it is well known that shorter proofs tend to involve higher proving times, reducing the prover's overhead at the cost of increased verification overhead seems favorable in the common situation in which the holder runs a digital wallet on a mobile phone whereas the relying is running a dedicated server to verify \acp{vp}.

Our implementation and experiments suggest that future \acp{vp} could benefit from a very broad spectrum of predicate proofs that can be implemented. Yet, to make this a reality, we need to standardize formats for constraint systems, witnesses, and proving and verification algorithms, so developers can combine different libraries that reflect the application domain and hardware they will run on in a modular way. 
With proof generation performance on mobile phones arguably being the key weakness of the \acp{snark}-based approach, novel tools for efficiently loading files associated with witness and proof generation, or for improving proving speed by leveraging GPUs~\citep{ni2022enabling}, are also critical. Even if \ac{snark} generation is not yet possible on all devices or some users do not want to wait longer than they are used to, there are different solutions available in the short term. 
For example, users with lower privacy requirements could use outsource proof generation to a single trusted yet randomly selected third party. Fortunately, there are now also solutions to outsource proof generation to several servers that collaboratively generate the \ac{snark} in a multi-party computation, such that all of them need to collude to compromise the user's privacy~\citep{ozdemir2022experimenting}. Recently, \citet{chiesa2023eos} also proposed a way to outsource the major share of proving effort to an untrusted third party; lowering end-to-end latency by more than an order of magnitude. This would reduce even the time for generating a \ac{snark} for the verification of an \ac{ecdsa} signature on a mobile phone to around one second. Specifically for less performant devices or exceptionally complex proofs, this approach may help achieving much shorter proof generation time and, therefore, quick broad adoption despite the presence of less performant mobile phones and the restriction of secure elements to non-\ac{snark}-friendly digital signature mechanisms.

Besides the apparent need for further performance improvements, in particular when \ac{sha256} and \ac{ecdsa} need to be involved, several other challenges outside the scope of \acp{snark} remain for data-minimal digital identity infrastructures that provide promising avenues for future research.
For instance, related work indicates that although users believe in privacy benefits of \ac{ssi}~\citep{guggenberger2023accept}, they struggle with understanding the new privacy capabilities that digital wallets and general-purpose \acp{zkp} offer~\citep{sartor2022ux}. This situation highlights the need for user acceptance studies involving the capabilities of sophisticated anonymous credentials. For instance, different visual cues may be used to illustrate the capabilities of \ac{snark}-based digital wallets to avoid the disclosure of different types of unique cryptographic identifiers or to generate predicate proofs in a way that makes end users appreciate them. Considering the limited performance of generating \acp{zkp} on a mobile phone, it would also be interesting to explore how
proof generation can be prepared in the background while users are inspecting the identity attributes or predicates to be released prior to giving their consent, and which waiting times users deem acceptable depending on their privacy preferences. 
Future security investigations could also consider which additional anonymization on other layers are required and accessible to users to avoid correlation beyond \acp{vp}, similar to discussions on anonymous digital payment systems such as~\citep{tinn2021cbdc,gross2021designing}, which point out the need for network-level anonymization, e.g., via onion routing~\citep{garrido2022revealing} through the Tor network~\citep{dingledine2004tor}.

Future research could also compare the practicality of the two core paradigms that general-purpose \acp{zkp} facilitate: The first approach involves retrofitting existing credential standards by transpiling the corresponding verification libraries, as proposed in \citet{delignat2016cinderella}'s approach for X.509 certificates, or by implementing novel verification libraries for such credential systems in one of the available \acp{dsl}, as has been done for JWTs~\citep{zklogin,snark2021jwt}. The second approach involves creating a novel infrastructure of certificates tailored towards multi-credential presentations and \ac{snark}-friendly cryptographic primitives, closer to what Hyperledger AnonCreds aim to achieve and what~\citet{maram2021CanDID,rosenberg2022zkcreds} and our work propose. For such novel frameworks, the heuristic approaches we discuss in this paper should be formalized, for instance, to get provable privacy guarantees of the core components of \acp{vp} also under many concurrent interactions as well as the more complex setup that needs to be considered for studying the role of how combining designated verifier presentations can leverage \acp{snark} and asymmetric encryption to address \ac{mitm} attacks without restrictive certification and identification of the verifier. Of course, hybrid approaches are also conceivable, similar to the opportunity to chain \ac{w3c} Verifiable Credentials to \ac{ssl} certificates via publishing the corresponding key-pairs on a website. However, such patchworks of multiple standards would arguably increase the complexity of creating the related wallets and \ac{snark} proving and verification backends.
Future research is also required to compare different approaches to revealing multiple attributes from several credentials in a \ac{vp}, which is common in practical applications. As we discussed in Section~\ref{sec:design}, the holder could generate multiple \acp{zkp} that reveal attributes from each credential, or a single \ac{zkp} that takes multiple credentials as input and verifies all of them. The corresponding trade-offs depend on parameters such as the number of credentials involved and on whether general-purpose \acp{zkp} with or without a trusted setup are used.

While privacy-focused digital identity management is particularly important for individuals, the application of \acp{snark}-based anonymous credentials is relevant far beyond. Digital identities for machines are also exposed to privacy issues because they are often associated with individuals. For instance, blockchain-based energy markets require fine-granular and, therefore, personally identifiable production and consumption information~\citep{roth2022blockchainenergy,babel2022enabling}. Verifiability can typically be provided by signing the data with certified sensors, yet information such as the exact location of the sensor may be too sensitive to reveal. In this context, advanced predicates like proofs of geo-location can be useful. Further applications where verifiable information for identification, authentication, and access control is desirable yet predicates may become relatively complex comprise blockchain-based e-voting~\citep{delignat2016cinderella}, privacy-focused digital currencies in regulated environments~\citep{gross2021designing,wust2022platypus}, access management~\citep{difrancescomeasea2023xacmlzkp}, verifiable polling that offers plausible deniability through local differential privacy~\citep{garrido2022polling}, combating disinformation~\citep{sedlmeir2023battling},
as well as various applications in the Metaverse~\citep{dwivedi2022metaverse}.

Privacy-focused and user-centric means of digital identification, authentication, and authorization verification have come a long way since~\citeauthor{chaum1985security}'s seminal paper. The pilot projects and political developments around \ac{ssi} that we can observe look promising and provide a unique opportunity to safeguard individuals' privacy. However, they still pose risks owing to the presence of cryptographic unique identifiers and an increasing amount of verifiable identity information that will arguably be exchanged in the future digital economy. Today's practical implementations of anonymous credentials that can address these problems still have significant shortcomings, as they build on hand-crafted and, thus, highly performant but inflexible \acp{zkp} designed in the early 2000s. This paper points to many areas where the limited functionality of these approaches causes tensions with requirements in regulated environments and existing identity and trust infrastructures. In the last few years, there has been impressive progress on the performance and ease of implementation of general-purpose \acp{zkp} like \acp{snark}, driven by (zk-)SNARK-based privacy and scaling solutions for cryptocurrencies. This makes a much more powerful technology stack available to implement privacy-focused \ac{ssi} systems to a much larger community of developers. By illustrating the ways in which general-purpose \acp{zkp} can address pressing problems of today's privacy-oriented implementations of \ac{ssi}, and that performance can be considered practical, this paper aims to encourage stakeholders on all levels to include general-purpose \acp{zkp} in their technical roadmap for privacy-oriented \ac{ssi} solutions and to invest in exploring the corresponding novel opportunities in the design space of digital identity infrastructures.

\subsection*{Acknowledgements}
We gratefully acknowledge the Bavarian Ministry of Economic Affairs, Regional Development and Energy for
its funding of the project ``Fraunhofer Blockchain Center'' (20-3066-2-6-14) and the
Luxembourg National Research Fund (FNR) for its support of the PABLO project (ref.~16326754). We also thank Iván Abellán Álvarez for his support with the \texttt{extractKthBit} method, Egor Ermolaev for his support in implementing the multi-threaded generation of auxiliary inputs for \ac{ecdsa} circuits, Pavel Zářecký for contributing the \ac{snark}-generation proving time measurements on two mobile phones, and Alexander Rieger for his valuable suggestions for improvement of the manuscript. For the purpose of open access, and in fulfilment of the obligations arising from the grant agreement, the authors have applied a Creative Commons Attribution~4.0 International (CC~BY~4.0) license to any Author Accepted Manuscript version arising from this submission.

\subsection*{CRediT author statement}

\textbf{Matthias Babel:} 
Conceptualization, Software, Validation, Investigation, Writing -- review and editing, Visualization. 

\textbf{Johannes Sedlmeir:}
Conceptualization, Software, Validation, Data curation, Investigation, Writing -- original draft, Visualization, Supervision.

\subsection*{Declaration of competing interests}
The authors declare that they have no known competing financial interests or personal relationships that could have appeared to influence the work reported in this paper.



%% References

%% References { BibTeX database:

%\bibliography{style{elsarticle-harv}

\clearpage

\bibliographystyle{elsarticle-num-names} %apalike
\bibliography{references}

\clearpage

%\printbibliography






% \end{thebibliography}

\clearpage
\appendix
\onecolumn
\section{Credential design}

\begin{figure}[H]
    \centering
    \resizebox{0.35\linewidth}{!}{%
    \begin{tikzpicture}
    \tikzset{every tree node/.style={align=center,anchor=north}}
    \Tree [.IssuerPK~\&~Signature
            [.Root 
                [. Meta~Root \edge node[auto=right,pos=0.5] {\color{orange}{Merkle tree}};
                    [. \node [below=0.4cm] {Metadata~leaves\\\emph{BigInt~(253~bits)}};
                    \edge node[auto=right,pos=0.5] {\color{orange}{encoding}};
                        [. \node [below=1cm] {Metadata \\\emph{String, Integer, Float, Boolean}};
                        ]
                    ]
                ]
                [. Content~Root \edge node[auto=right,pos=0.5] {\color{orange}{Merkle tree}};
                    [. \node [below=0.4cm] {Attribute~leaves\\\emph{BigInt~(253~bits)}};
                    \edge node[auto=right,pos=0.5] {\color{orange}{encoding}};
                        [. \node [below=1cm] {Attributes \\\emph{String, Integer, Float, Boolean}};
                        ]
                    ]
                ]
            ]
          ]
    \end{tikzpicture}
    }
    \vspace{1em}
    \subcaption{\label{fig:credential} Overall credential form.}
    \medskip
    \vspace{30pt}
    \centering
        \resizebox{0.55\linewidth}{!}{%
        \begin{tikzpicture}
        \tikzset{every tree node/.style={align=center,anchor=north}}
        \Tree   [.Meta\ Root 
                    [.H0123 
                        [.H01 
                            [.H0 Revoc.ID ]
                            [.H1 Schema ]
                        ] 
                        [.H23 
                            [.H2 BindingPK.x ]
                            [.H3 BindingPK.y ]
                        ] 
                    ] 
                    [.H4567 
                        [.H45 
                            [.H4 Revoc.~Reg. ]
                            [.H5 Exp.~Date ]
                        ] 
                        [.H67 
                            [.H6 Delegatable? ]
                            [.H7 Empty ]
                        ] 
                    ] 
                ]
        \end{tikzpicture}
        }
    \vspace{1em}
    \subcaption{\label{fig:meta_tree} Metadata Merkle tree example.}
    \medskip
    \vspace{30pt}
    \centering
        \resizebox{0.45\linewidth}{!}{%
        \begin{tikzpicture}
        \tikzset{every tree node/.style={align=center,anchor=north}}
        \Tree   [.Content\ Root 
                    [.H0123 
                        [.H01 
                            [.H0 Name ]
                            [.H1 Surname ]
                        ] 
                        [.H23 
                            [.H2 Gender ]
                            [.H3 Birthdate ]
                        ] 
                    ] 
                    [.H4567 
                        [.H45 
                            [.H4 Eye~color ]
                            [.H5 Height ]
                        ] 
                        [.H67 
                            [.H6 Address.x ]
                            [.H7 Address.y ]
                        ] 
                    ] 
                ]
        \end{tikzpicture}
        }
    \vspace{1em}
    \subcaption{\label{fig:content_tree} Content Merkle tree example.}
\end{figure}

\clearpage

\section{Selected Code Snippets}
\FloatBarrier

The following code snippets aim to illustrate how to implement verifiable computations beyond simple quadratic operations in Circom. Essentially, the operations required in our data-miminizing \ac{vp} implementation are hashing (Merkle trees and Merkle proofs, occurs in many places), digital signature verifications (integrity and holder binding), range proofs (non-expiration, age proof), and integer division, modulo operation, and extraction of the k$^{\mathrm{th}}$ bit of a given number (Merkle tree-based revocation). Poseidon (and also~\ac{sha256}) hashing and digital signature verification (\ac{eddsa}) are already provided in the Circomlib library~\citep{iden32022circomlib}, and circuits for re-computing Merkle roots from Merkle trees and Merkle proofs are a core part of almost every blockchain-related implementation of \acp{snark} and implemented in many examples~\citep[e.g.,~][]{github2022rollup}. The provided snippets therefore almost completely represent the remaining more complex generic components needed for our implementation of standard \ac{snark}-based data-minimal verifiable presentations in scenarios (I) to (VII). We also demonstrate how to implement the polygon inbound proof as an example of a complex predicate.

\subsection{Simple basic operations with Circom}
\begin{figure}[!h]
%\centering
    \begin{subfigure}[t]{.01\linewidth}
    ~
    \end{subfigure}
    \begin{subfigure}[t]{.2\linewidth}
        \begin{tabular}{c}
            %\resizebox{1.1\linewidth}{!}{ 
\begin{lstlisting}[language=Circom, basicstyle=\tiny]
pragma circom 2.1.0;

template Cube() {
  signal input a;

  signal b;
  signal output c;

  b <== a * a;
  c <== b * a;

}
\end{lstlisting}
        %}
        \end{tabular}
        \caption{Example for raising to the power of three in Circom, i.e., proving that a private input is a cubic root.}
        \label{code:cubic}
    \end{subfigure}
    \begin{subfigure}[t]{.1\linewidth}
    ~
    \end{subfigure}
    \begin{subfigure}[t]{.287\linewidth}
        \begin{tabular}{c}
        %\resizebox{0.95\linewidth}{!}{
\begin{lstlisting}[language=Circom,  
        basicstyle=\tiny]
pragma circom 2.1.0;

include "poseidon.circom";

template myHasher() {
    
    signal input x;
    signal input y;

    signal output z;

    component hasher = Poseidon(2);
    hasher.inputs[0] <== x;
    hasher.inputs[1] <== y;

    z <== hasher.out;

}
\end{lstlisting}
        %}
        \end{tabular}
        \caption{Example for implementing hashing in Circom, i.e., proving that two private child leaves yield a given public Poseidon hash.}
    \label{code:pre-image}
    \end{subfigure}
\end{figure}

\subsection{Integer division and modulo operations}
\begin{figure*}[!h]
    %\begin{tabular}{cc}
    %\phantom{h} & \resizebox{1.5\linewidth}{!}{
\begin{lstlisting}[language=circom, basicstyle=\tiny, escapeinside={@}{@}]
pragma circom 2.1.0;

include "comparators.circom";

template IntegerDivisionWithRest(n) {

    // Input signals
    signal input in; // The number to be divided
    signal input div; // The number by which is divided. Must be smaller than @\color{gray}$2^n$@.
    
    // Output signals
    signal output multiplier; // The result of the integer division ("in \ div")
    signal output rest; // The rest of the integer division ("in % div")

    // Make sure that the integer is in the prime field.
    assert(n <= 253); 

    // Used components
    // Range check imported from comparators.circom that takes two inputs and returns 1 if the first input is smaller than the second and 0, else.
    component rangeCheck = LessThan(n);

    // Supply the result of integer division and the rest "magically" (avoids bloating the program with additional (private) inputs or doing more complex computations than necessary inside the circuit).
    multiplier <-- in \ div; // Integer division, no constraints assigned.
    rest <-- in % div; // Modulo operation, no constraints assigned.
    
    // Check that this is the unique correct result
    in === multiplier * div + rest; // (1) input must be recovered from multiplier, div, and result.
    rangeCheck.in[0] <== rest;
    rangeCheck.in[1] <== div;
    rangeCheck.out === 1; // (2) rest must be smaller than div.

}
\end{lstlisting}
    %}
    %\end{tabular}
    \caption{Circom implementation of a circuit for  integer division with rest. The rest can be used for obtaining the result of a modulo operation.}
    \label{code:modulo}
\end{figure*}

\clearpage
\subsection{Extracting a specific bit from a number}
\begin{figure*}[!h]
    %\begin{tabular}{cc}
    %\phantom{h} & \resizebox{1.5\linewidth}{!}{
\begin{lstlisting}[language=circom, basicstyle=\tiny, escapeinside={@}{@}]
pragma circom 2.1.0;

include "comparators.circom";

template extractKthBit(n) {
    /*
    Given a 253-bit BigInt as input, this template computes the binary representation and returns the corresponding bit at the k-th position (starting the counting from 1 at the least significant bit!). 
    For instance, 199 @\color{gray}$\mathrel{\stackon[1.5pt]{=}{\stretchto{\scalerel*[\widthof{=}]{\wedge}{\rule{1ex}{3ex}}}{0.5ex}}}$@ [0, @\ldots@, 0, 1, 1, 0, 0, 0, 1, 1, 1], where k = 3 -> output = 1, for k = 4 -> output = 0.
    */

    // Make sure that the integer is in the prime field.
    assert(n <= 253);
    
    signal input in;
    signal input k;

    signal binaryRepresentation[n]; // Binary representation of in, used as intermediate signal
    
    var powersOfTwo = 1; // Used to iteratively compute powers of two
    var runningBinarySum = 0; // Used to recover in from its binary representation (necessary as the decomposition to binary is not constrained).
    signal runningOutputBitSum[n+1]; // Used to iteratively compute @\color{gray}$\sum_{i=0}^n$@ binaryRepresentation[i] * IsEqual(i,k), where IsEqual() outputs 1 if True, 0 if False.
    runningOutputBitSum[0] <== 0;
    
    component equalityCheck[n]; // Used for populating with many IsEqual() components, one for every loop.

    for (var i = 0; i<n; i++) {
    
        binaryRepresentation[i] <-- (in >> i) & 1; // Extract the bit value through a (non-verified) Oracle call.
        binaryRepresentation[i] * (binaryRepresentation[i]-1) === 0; // Ensure that the bit is indeed boolean, i.e., 0 or 1.

        equalityCheck[i] = IsEqual(); // Initialize an IsEqual() component from comparators.circom.
        equalityCheck[i].in[0] <== k-1; // First input of the equality comparison.
        equalityCheck[i].in[1] <== i; // Second input of the equality comparison.
        // If i = k, add this bit to runningOutputBitSum; else, there should be no contribution.
        runningOutputBitSum[i+1] <== runningOutputBitSum[i] + binaryRepresentation[i] * equalityCheck[i].out;
        
        // Iteratively recover the input from its supposed binary representation.
        runningBinarySum += binaryRepresentation[i] * powersOfTwo;
        powersOfTwo += powersOfTwo;
    }

    // Make the remaining check to verify that the Oracle-provided binary representation is legitimate. Inspired by the Num2Bits implementation 
        (@\color{blue}  \url{https://github.com/iden3/circomlib/blob/master/circuits/bitify.circom}@).
    runningBinarySum === in;
    
    // Assign the output bit by using the last value of the runningOutputBitSum.
    signal output outBit;
    outBit <== runningOutputBitSum[n];
}
\end{lstlisting}
    %}
    %\end{tabular}
    \caption{Circom implementation of a circuit for retrieving the \texttt{k}$^{\mathrm{th}}$ bit from a large integer \texttt{input} with at most 253~bits. % (i.e., the maximum number that can be used as input is 14474011154664524427946373126085988481658748083205070504932198000989141204991).
    Note that the subsequent values of \texttt{powersOfTwo} in each loop cycle do not need to be constrained because they are known at compile time, which is why we can make it a \texttt{var} instead of a \texttt{signal}. The subsequent values for \texttt{runningBinarySum} also only need to be constrained in the end (line~42) once. This is because \texttt{runningBinarySum} is only a linear combination (with coefficients from \texttt{powersOfTwo} that are known at compile time) of signals, namely the elements of the \texttt{binaryRepresentation} array. In contrast, every update of the \texttt{runningOutputBitSum} is non-linear, so there must be new assignments (in an array of signals) in each individual step (see line~35).}
    \label{code:extractKthBit}
\end{figure*}

\clearpage
\subsection{Polygon in-/outbound proof (``leather-pants predicate'')}
\begin{figure*}[!h]
    \centering
    %\begin{tabular}{cc}
    %\phantom{h} & \resizebox{2.3\linewidth}{!}{
\begin{lstlisting}[language=Circom,       basicstyle=\tiny, escapeinside={@}{@}]
pragma circom 2.1.0;

include "comparators.circom";
include "modulo.circom";

template Polygon(n) {

  signal input vertx[n]; // x-coordinates of vertices
  signal input verty[n]; // y-coordinates of vertices
  signal input x; // x-coordinate
  signal input y; // y-coordinate
  signal output out; // 1 if inside, 0 else

  var j = n-1;
    
  signal c[n+1]; // Counter for intersections.
  c[0] <== 0;
    
  signal tmp[5][n];
  signal bool[3][n];
	
  // Preparing for some comparator components.
  component gt[4][n];
  component st[n];

  for (var i=0; i<n; i++) {
	
    // Checking if y is between the y-coordinates of the two vertices.
    gt[0][i] = GreaterThan(64);
    gt[0][i].in[0] <== verty[i];
    gt[0][i].in[1] <== y;
    gt[1][i] = GreaterThan(64);
    gt[1][i].in[0] <== verty[j];
    gt[1][i].in[1] <== y;
    tmp[0][i] <== (1 - gt[0][i].out) * gt[1][i].out; // 1     iff verty[i] < y < verty[j], 0 else.
    tmp[1][i] <== gt[0][i].out * (1 - gt[1][i].out); // 1 iff verty[j] < y < verty[i], 0 else.
		
    bool[0][i] <== tmp[0][i] + tmp[1][i]; // 1 iff y is between the y coordinates of vertex i and vertex j, 0 else.
		
    // Checking which of the y coordinates of the two vertices is larger.
    tmp[2][i] <== verty[j] - verty[i];
    gt[2][i] = GreaterThan(64);
    gt[2][i].in[0] <== tmp[2][i];
    gt[2][i].in[1] <== 0;
		
    // Checking whether the connecting line between the vertices i and j intersects the horizontal rail to the right (see Franklin (2006)). We multiply x - vertx[i] @$\color{gray}{<}$@ (vertx[j] - vertx[i]) / (verty[j] - verty[i]) * (y - verty[i]) with the denominator and check the two cases separately to account for the sign.
    tmp[3][i] <== (x - vertx[i]) * tmp[2][i];
    tmp[4][i] <== (vertx[j] - vertx[i]) * (y - verty[i]);
		
    gt[3][i] = GreaterThan(64);
    gt[3][i].in[0] <== tmp[3][i];
    gt[3][i].in[1] <== tmp[4][i];
    st[i] = LessThan(64);
    st[i].in[0] <== tmp[3][i];
    st[i].in[1] <== tmp[4][i];
    bool[1][i] <== gt[2][i].out * gt[3][i].out;
    bool[2][i] <== (1 - gt[2][i].out) * st[i].out;
		
    // If one of the two cases holds, increase c by one.
    c[i+1] <== c[i] + bool[0][i] * (bool[1][i] + bool[2][i]);
		
    // Make sure that j,i are two subsequent vertex indices in every loop, cycling the whole polygon.
    j = i;

  }
    
  // Checking if out is odd via the modulo component (see also Figure @\ref{code:modulo}@).
  component modulo = IntegerDivisionWithRest(n);
  modulo.in <== c[n];
  modulo.div <== 2;
	
  out <== modulo.rest; // Returning 1 (true) if c was odd, i.e., (x,y) is inside the polgyon. 

}
\end{lstlisting}
    %}
    %\end{tabular}
    \caption{Circom implementation of a circuit that can be used to create a \ac{snark} that a given point (x,y) in the Euclidean plane is inside or outside a given polgyon with~n~vertices. When using this component in a \ac{vp}, we consider the case where (x,y) is private (with the root of trust being the signature of the issuer on a credential that includes the coordinates as attributes) while the polygon is public (i.e., the coordinates of all vertices are specified by the verifier in the proof request).}
    \label{code:polygon}
\end{figure*}

\end{document}